\documentstyle[12pt,epsf]{article}

\begin{document}
\title
{
Discrete structure of ultrathin dielectric films and their surface
optical properties
}

\author
{
S.V.Sukhov$^1$ and K.V.Krutitsky$^{1,2}$\\
$^1$Ulyanovsk Branch of Moscow Institute of Radio Engineering\\
and Electronics of Russian Academy of Sciences,\\
P.B.9868, 48, Goncharov Str., Ulyanovsk 432011, Russia\\
e-mail: ufire@mv.ru
\\
$^2$Fakult\"at f\"ur Physik,
Universit\"at Konstanz, Fach M 674,\\
D-78457 Konstanz, Germany\\
e-mail: kostya@spock.physik.uni-konstanz.de
}

\date{\today}

\maketitle

\begin{abstract}
The boundary problem of linear classical optics about the interaction
of electromagnetic radiation with a thin dielectric film has been solved
under explicit consideration of its discrete structure.
The main attention has been paid to the investigation of the near-zone
optical response of dielectrics.
The laws of reflection and refraction for discrete structures
in the case of a regular atomic distribution are studied and
the structure of evanescent harmonics induced by an external plane wave
near the surface is investigated in details.
It is shown by means of analytical and numerical calculations
that due to the existence of the evanescent harmonics the laws of reflection
and refraction at the distances from the surface less than two
interatomic distances are principally different from the Fresnel
laws.
From the practical point of view the results of this work might be
useful for the near-field optical microscopy of ultrahigh resolution.
\end{abstract}

\begin{flushleft}
PACS numbers: 78.20.-e, 81.40.Tv, 42.25.-p, 87.64.Xx
\end{flushleft}

\section{Introduction}

Usually in the investigations of optical phenomena in dielectrics
a medium is considered as a continuous system of radiators.
Besides the macroscopic Maxwell equations are employed.
This way of description works well if one considers the phenomena
which take place at large enough distances from medium
interfaces.
However, in recent years the interest to the optical phenomena
near the surfaces of different media at the distances much less than
the radiation wavelength has dramatically increased.
Beside a purely fundamental interest this is connected to the
vigorous development of the scanning near-field optical
microscopy~\cite{NFO}-\cite{Vahid}.
There were developed various schemes of optical near-field
microscopes and the work in this direction is still currently
underway. However, all the schemes have a common feature, namely,
the measurement of the optical
response of the medium on the external radiation is carried out
in the near zone at the distances of the order of surface
inhomogenity~\cite{Xiao,Vahid}.

The resolution obtained with the aid of near-field optical microscopes
permanently increases and at the moment it is of the order
of 10 nm. Besides the probe scans the surface also at the distance of the
order of 10 nm~\cite{Vahid}.
At such a small distance it becomes necessary to take into account
the discrete structure of the medium.
The attempts of consideration of discrete medium structure in the
theoretical description of the near-field optical measurements have
been undertaken in recent years~\cite{Xiao}.
However, the main attention has been paid to the situations
when discrete dipoles model rather large surface inhomogenities
and the spatial distribution of these dipoles is asymmetric.
The progress in the near-field optical microscopy
allows one to hope that there will be possible to observe the atomic
structure of surfaces. Then it will be necessary to interpret
the optical response from single atoms more or less
regularly distributed on the surfaces. Similar problems occur
when one wants to describe the near-field in the vicinity of
the surfaces of photon~\cite{Miyazaki}
and plasmon~\cite{Smolyaninov} crystals.
Besides it becomes more natural to employ microscopic equations.

Intensive studies on optics of discrete dielectric media have been
started not very long time ago. The laws of refraction and reflection
for discrete structures were shown to be different from Fresnel
laws~\cite{KSJPB} and this leads to the anomalies in the behavior
of the electromagnetic field near the surface~\cite{KS}-\cite{PW}.
Theorem of extinction for discrete structures has been
derived and the corrections to the Snell law were worked out~\cite{Litzman}.
The characteristic features of the spontaneous emission
process~\cite{GKJETP,KA},
superradiation~\cite{ZMT} and optical bistability~\cite{MGF}
in discrete structures were investigated.
The second harmonic generation in discrete structures has been considered
in Ref.~\cite{WBHR}, where a good agreement with experimental data
has been pointed out. It was also shown that the consideration of
the discrete structure of the medium leads to the
surface local-field effect~\cite{WSM}
and surface induced optical anisotropy~\cite{MB}-\cite{Wijers}.

Properties of the waves reflected from discrete structures in the wave
and near zones have been studied as well. It has been shown that in
the wave zone the reflected waves can be described by plane
waves~\cite{KSJPB,KS,JS89,JSS,JS90}. However, the treatment
in Refs.~\cite{JS89,JSS,JS90} does not take into account evanescent
harmonics which give a considerable contribution into optical
response of discrete dielectric structures in the near zone.
These evanescent harmonics are not related to the total internal
reflection and they exist even in the case of a single atom~\cite{SKF}.
Due to the neglect of the evanescent harmonics the method developed
in Refs.~\cite{JS89,JSS,JS90} can be employed for the description of
the optical response of a thick enough medium in the wave zone only.

A more general approach which implicitly takes into account the evanescent harmonics
has been developed in Refs.~\cite{KSJPB,KS}. This allowed us to show that
the fields of the reflected waves in the near zone are periodic
functions of longitudinal coordinates with the period equal to the
lattice constant. This result confirms the assumption put forward
in Ref.~\cite{Vahid} that with the aid of near-field optical microscopes
one can observe the atomic structure of the medium surface.
In the present paper we shall continue the investigations started
in Refs.~\cite{KSJPB,KS}. Employing the methods of the calculation
of the lattice sums developed in Refs.~\cite{PWS,Wijers} we are going
to investigate in more details the behavior of evanescent harmonics near
the surface of a dielectric film with an ideal crystal structure.
We would like to note that the role of the evanescent harmonics in
the near-zone optical response of dielectrics has been point out
by Sivukhin~\cite{Siv85}. On the basis of rather simple and intuitive
physical arguments he came to the qualitative conclusion that
the Fresnel laws of reflection and refraction have to be violated at
the distances from the surface less than one interatomic distance.
In the present paper we are going to undertake a quantitative analysis
of this fundamental issue.
Besides it will be shown that the existence of the evanescent
harmonics leads to principal deviations from Fresnel laws of reflection and
refraction at the distances from the surface of the order of few
interatomic distances.

The paper is organized as follows. In section 2 we shall
discuss the problem we are going to solve and the main equations
which constitute the basis of the microscopic approach to
the description of optical phenomena in the linear classical optics.
Section 3 is devoted to the analyses of the probe influence on the
electromagnetic field distribution in the film and the back
influence of the perturbation caused by the probe on the field
magnitude on it.
It is shown that in some cases the probe influence can be neglected.
In section 4 we shall calculate the distribution of the local field
in a dielectric film under the incidence of a plane wave.
In section 5 we shall analyze in details the characteristic features
in the behavior of reflected and transmitted waves in the wave and
near zones.

\section
{
Statement of the problem and basic equations
}

Let a monochromatic light wave with frequency $\omega$ and electric
field strength vector
$
  {\bf E}_I({\bf r},t)
  =
  {\bf A}_I({\bf r})
  \exp
  \left(
        - i \omega t
  \right)
$
be incident on a dielectric film with thickness $h$.
Above the film there is a probe which measures the electromagnetic
field at a certain point of space.
The probe will be treated as a single point-like dipole with
a linear polarizability $\alpha_p$. Such a scheme can be realized
in a real experiment using either a single atom in a magneto-optical
trap~\cite{Kawata} or an admixed atom implanted in the needle of
a near-field microscope~\cite{Vahid, Letohov}.
Our aim is to investigate the behavior of the electromagnetic field
in the film and to calculate the characteristics of the electromagnetic
field on the probe that is moved (scanned) along the film surface
at some fixed distance $d$. The field detection, which
represents itself a separate problem, will not be touched in the
present paper.

We shall treat the dielectric film as a system of $N$ monolayers
located at a distance $a$ from one another. The atoms inside
the monolayers possess a discrete spatial distribution.
In a stationary case the equation for the local electric field
strength on the probe
  $
  {\bf E}_p'({\bf r},t)
  =
  {\bf E}_p({\bf r})
  \exp
  \left(
        - i \omega t
  \right)
$
\cite{Rosenfeld,BW}
can be written down in the form
\begin{equation}
\label{field-p}
{\bf E}_p({\bf r})
=
{\bf A}_I({\bf r})
+
i \frac{2}{3} k_0^3 \alpha_p {\bf E}_p({\bf r})
+
\alpha
\sum_{j=1}^N
\sum_{a_j}
     \nabla_{{\bf r}} \times \nabla_{{\bf r}} \times
     {\bf E}_j({\bf r}_{a_j}^\|)
     G(R_{a_j})
,
\end{equation}
where ${\bf r}=(x,y,z)$ is a radius-vector of the probe position;
$\alpha$ is the microscopic polarizability of the atoms inside the film,
which is assumed to be independent of the field.
$
 G(R)
 =
 \exp
 \left(
      i k_0 R
 \right)
 /R
$
is the Green function of Helmholtz equation, $k_0=\omega/c$
is the vacuum wave number,
${\bf R}_{a_j}={\bf r} - {\bf r}_{a_j}$,
$R_{a_j}=\left|{\bf R}_{a_j}\right|$,
${\bf r}_{a_j}=({\bf r}_{a_j}^\|, z_j)$.
We have also introduced the notation
\begin{equation}
{\bf E}_j({\bf r}_{a_j}^\|)
=
{\bf E}({\bf r}_{a_j}^\|,z_j),
\qquad z_j=-(j-1)a
,
\qquad
j=\overline{1,N}.
\end{equation}
The reference frame is chosen in such a manner that $z$ axis
is perpendicular to the film surface.
The second term on the r.h.s. of eq.(\ref{field-p}) describes
the Lorentz radiation damping~\cite{Rosenfeld,LL}.
In spite of the fact that the radiation damping makes usually
a small contribution into the resultant field the consideration
of this term is of principal significance. If one does not take
it into account the energy conservation is violated
(see~\cite{GS-97} and references therein).

Let
$
 {\bf r}_{a_l}=
 \left(
      x_{a_l} , y_{a_l} , z_l
 \right)
$
be a radius-vector of some atom in the $l$-th monolayer.
Then for the local field at the position of this atom we have
\begin{eqnarray}
\label{field-m}
{\bf E}_l
\left(
     {\bf r}_{a_l}^\|
\right)
&=&
{\bf E}_{Il}
\left(
     {\bf r}_{a_l}^\|
\right)
+
i \frac{2}{3} k_0^3 \alpha
{\bf E}_l
\left(
     {\bf r}_{a_l}^\|
\right)
+
\alpha_p
\nabla_{
        {\bf r}_{a_l}
       }
\times
\nabla_{
        {\bf r}_{a_l}
       }
\times
{\bf E}_p({\bf r})G(R_{a_l})
\nonumber\\
&+&
\alpha
\sum_{j=1}^N
\sum_{a_j}
\nabla_{
        {\bf r}_{a_l}
       }
\times
\nabla_{
        {\bf r}_{a_l}
       }
\times
{\bf E}_j({\bf r}_{a_j}^\|)
G(R_{a_l a_j})
,
\qquad
l=\overline{1,N},
\end{eqnarray}
where
$
{\bf E}_{Il}
\left(
     {\bf r}_{a_l}^\|
\right)
=
{\bf A}_I
\left(
     {\bf r}_{a_l}
\right)
$,
$
R_{a_l a_j}=
\left|
     {\bf r}_{a_l} - {\bf r}_{a_j}
\right|
$.
Note that the term with ${\bf r}_{a_j} = {\bf r}_{a_l}$ has to be
excluded from the summation.

We can look for the solution in the $l$-th monolayer of the film
in the form
\begin{equation}
\label{E-l0p}
{\bf E}_l
\left(
     {\bf r}_{a_l}^\|
\right)
=
{\bf E}_l^0
\left(
     {\bf r}_{a_l}^\|
\right)
+
{\bf E}_l^p
\left(
     {\bf r}_{a_l}^\|
\right)
,
\end{equation}
and in addition we impose a requirement that the field
$
{\bf E}_l^0
\left(
     {\bf r}_{a_l}^\|
\right)
$
has to satisfy the equation
\begin{equation}
\label{E-0}
{\bf E}_l^0
\left(
     {\bf r}_{a_l}^\|
\right)
=
{\bf E}_{Il}
\left(
     {\bf r}_{a_l}^\|
\right)
+
i \frac{2}{3} k_0^3 \alpha
{\bf E}_l^0
\left(
     {\bf r}_{a_l}^\|
\right)
+
\alpha
\sum_{j=1}^N
\sum_{a_j}
\nabla_{
        {\bf r}_{a_l}
       }
\times
\nabla_{
        {\bf r}_{a_l}
       }
\times
{\bf E}_j^0({\bf r}_{a_j}^\|)
G(R_{a_l a_j})
.
\end{equation}
Then from eqs.(\ref{field-p}), (\ref{field-m})
it follows that the fields
${\bf E}_l^p$ and ${\bf E}_p$
will be the solutions of the system of equations
\begin{eqnarray}
\label{E-probe}
{\bf E}_p({\bf r})
=
{\bf A}_I({\bf r})
+
i \frac{2}{3} k_0^3 \alpha_p
{\bf E}_p({\bf r})
&+&
\alpha
\sum_{j=1}^N
\sum_{a_j}
     \nabla_{{\bf r}} \times \nabla_{{\bf r}} \times
     {\bf E}_j^0({\bf r}_{a_j}^\|)
     G(R_{a_j})
\nonumber\\
&+&
\alpha
\sum_{j=1}^N
\sum_{a_j}
     \nabla_{{\bf r}} \times \nabla_{{\bf r}} \times
     {\bf E}_j^p({\bf r}_{a_j}^\|)
     G(R_{a_j})
,
\end{eqnarray}
\begin{eqnarray}
\label{E-p}
{\bf E}_l^p
\left(
     {\bf r}_{a_l}^\|
\right)
=
i \frac{2}{3} k_0^3 \alpha
{\bf E}_l^p
\left(
     {\bf r}_{a_l}^\|
\right)
&+&
\alpha_p
\nabla_{
        {\bf r}_{a_l}
       }
\times
\nabla_{
        {\bf r}_{a_l}
       }
\times
{\bf E}_p({\bf r})G(R_{a_l})
\nonumber\\
&+&
\alpha
\sum_{j=1}^N
\sum_{a_j}
\nabla_{
        {\bf r}_{a_l}
       }
\times
\nabla_{
        {\bf r}_{a_l}
       }
\times
{\bf E}_j^p({\bf r}_{a_j}^\|)
G(R_{a_l a_j})
.
\end{eqnarray}
The physical meaning of the fields
${\bf E}_l^0$ and ${\bf E}_l^p$ is the following.
${\bf E}_l^0$ is the local field inside the film without consideration
of the contribution made by the probe. This contribution is taken
into account by means of the field ${\bf E}_l^p$.
Thus, for the solution of the problem stated above it is necessary
first to solve the unperturbed problem (\ref{E-0}). Then using
the obtained solution one can determine the perturbation of the field
inside the film caused by the probe and calculate the back influence of
this perturbation on the field at the position of the probe
(\ref{E-probe}), (\ref{E-p}).
However, before dealing with this general problem we shall estimate
the perturbation in the film caused by the probe and find out under which
conditions it is considerable.

\section
{
Estimate of the perturbation caused by the probe
}

In this section we shall estimate ${\bf E}_l^p$ which enters
eqs.(\ref{E-probe}), (\ref{E-p}). For that purpose one can neglect
radiation damping terms in eqs.(\ref{E-probe}), (\ref{E-p}).
Our starting point is the following.
The perturbation of the field in the medium
${\bf E}_l^p$, which is caused by the probe,
decreases rapidly with the increase of the distance between the surface
and the probe. If in addition we take into account that the influence
of atoms on one another also rapidly decreases with the distance, one
can come to the conclusion that either the probe or the nearby atoms
will be influenced mainly by the atoms located in the nearest vicinity
around them. Thus, it is necessary to find the field
${\bf E}^p ({\bf r}_{a})$ in the direct vicinity of the probe.
Let's estimate first the size of this region.

One can assume that the behavior of the field ${\bf E}_l^p$
has the same features as that induced by the probe
\begin{eqnarray}
\label{29}
{\bf E}_l^p ({\bf r}_{a_l}^\|) & \propto &
\left[
\left(
\frac{3({\bf E}_p {\bf R}_{a_l}){\bf R}_{a_l}}{R_{a_l}^5}
-
\frac{{\bf E}_p}{R_{a_l}^3}
\right)
-
i k_0
\left(
\frac{3({\bf E}_p {\bf R}_{a_l}){\bf R}_{a_l}}{R_{a_l}^4}
-
\frac{{\bf E}_p}{R_{a_l}^2}
\right)
-
\right.
\\ \nonumber
& &
\left.
k_0^2
\left(
\frac{({\bf E}_p {\bf R}_{a_l}){\bf R}_{a_l}}{R_{a_l}^3}
-
\frac{{\bf E}_p}{R_{a_l}}
\right)
\right]
\exp(i k_0 R_{a_l})
.
\end{eqnarray}
Taking into account that the field on the probe is defined by
the atoms in the near zone, one can keep in eq.(\ref{29})
only the terms proportional to $R_{a_l}^{-3}$. We substitute
the corresponding expression into (\ref{E-p}) and replace the
summation by integration. As a result of this the expression
for the third term in the r.h.s. of (\ref{E-p}) takes the form
\begin{equation}
\label{29a}
\alpha
\sum_{j=1}^N
\sum_{a_j}
\nabla_{
        {\bf r}_{a_l}
       }
\times
\nabla_{
        {\bf r}_{a_l}
       }
\times
{\bf E}_j^p({\bf r}_{a_j}^\|)
G(R_{a_l a_j})
\to
\rho
\int_{\Sigma_\sigma}^{\Sigma}
\nabla_{
        {\bf r}
       }
\times
\nabla_{
        {\bf r}
       }
\times
{\bf E}^p({\bf r}')d{\bf r}'
,
\end{equation}
with
$\rho$ being the macroscopic concentration of atoms. The integration in (\ref{29a})
is carried out over the film volume limited by the external surface
$\Sigma$. In order to prevent the interaction of the atom with itself
a small region limited by the spherical surface $\Sigma_\sigma$
centered at the observation point has to be excluded from the integration.
The size $\sigma$ of this small region is of the order of lattice
constant. In Ref.~\cite{KSJPB} it was shown that in order to achieve
the complete correspondence passing from discrete to continuous atomic
distribution one has to put $\sigma=\frac34 a$.

Let's determine the region of space, atoms from which make the main
contribution to the field on the atom nearest to to the probe. Let
$\delta$ be a tolerance of the calculations of the dipole field. Then
  \begin{equation}
\label{est}
1
-
\int_{\Sigma_\sigma}^{\Sigma_L}
\nabla_{
        {\bf r}
       }
\times
\nabla_{
        {\bf r}
       }
\times
{\bf E}^p({\bf r}')d{\bf r}'
\left/
\int_{\Sigma_\sigma}^{\Sigma}
\nabla_{
        {\bf r}
       }
\times
\nabla_{
        {\bf r}
       }
\times
{\bf E}^p({\bf r}')d{\bf r}'
\right.
=
\delta
,
\end{equation}
Integration in the denominator and in the numerator in eq.(\ref{est})
is carried out over the film volume and over the spherical segment
with the radius $L$ centered at the position of the atom nearest to
the probe, respectively.
Taking into account (\ref{29}) we obtain from eq.(\ref{est})
the following estimate for the radius of the sphere $L$ within which
it is necessary to calculate the fields on the atoms:
\begin{equation}
L\approx \sigma\delta^{-1/3}.
\end{equation}
Let
$\delta =0.1\%$, then $L\approx 7a$.
Note that in the calculations one can use even less values of $L$,
because in eq.(\ref{E-p})
there is also the field induced by the probe and compared to this
field the influence of the remote atoms is less. The conclusion that
only the neighboring atoms contribute to the field
at the site of probe in near-field regime
is in agreement with the results of Ref.~\cite{Keller},
where the influence on the probe caused by microscopic dielectric
spheres located at the surface has been calculated. It was shown
that the lattice of $5\times 5$ spheres makes the main contribution.
The increase of this area does not influence the final result.

Performing analogous calculations for eq.(\ref{E-probe})
we obtain that the main contribution to the field on the probe
is made by the atoms located inside the sphere with the radius
$L$ centered at the probe position. Thus, the system of equations
(\ref{E-probe}), (\ref{E-p}) is reduced to the system of few
linear algebraic equations, which can be solved numerically
after the field ${\bf E}^0_l$ is obtained from eq.(\ref{E-0}).
However, as it will be shown below in many situations it is even
not necessary to solve this system of equations due to the negligible
contribution of ${\bf E}^p$.

For the sake of distinctness we assume the probe to be located above
some surface atom of the film. The field induced by the probe at the
position of this atom is determined by the expression
\begin{equation}
\label{6}
{\bf E}_{eff}
\approx
-\alpha_p {\bf E}_p/d^3
,
\end{equation}
where $d$ is the distance between the probe and the atom.
In formula (\ref{6}) we have neglected the retardation and
the field on the probe is chosen to be polarized
parallel to the film surface assuming that the variations of
the direction of the polarization vector does not influence much the
final results. It has been shown above that in the calculations of
the field on the probe it is enough to take into account the atoms
located inside the sphere with the radius $L$.
The rest of the film does not influence much the probe.
So we shall estimate the influence on the probe, caused by the atoms
located within the spherical segment centered at the position of
the atom below the probe. Obviously, the microscopic field amplitude
on the atom under consideration is larger than that on the other atoms
of the film.
The upper estimate of the field amplitude at the position of this
atom ${\bf E}_{max}$ is given by
\begin{equation}
\label{30}
|{\bf E}_{max}|
\le
|{\bf E}_{eff}|
\left[
   1-1.71 C \ln(L/\sigma)
\right]^{-1}
,
\end{equation}
where the dimensionless parameter $C=\alpha\rho$ is a volumic
polarizability~\cite{JS90}.
We shall assume that the field on any atom inside the spherical
segment is constant and equal to the field on the central atom
${\bf E}_{max}$.
It is obvious that in this approximation the influence on the probe
caused by the spherical segment is larger than it is. Our calculations
are approximate, so we replace the summation by the integration.
Calculating the field at the position of the probe we get an upper
estimate of the last term in the r.h.s. of eq.(\ref{E-probe})
\begin{equation}
\label{31}
|{\bf E}_R^{eff}|
<
1.71 C |{\bf E}_{max}|
\ln
\left( 1+L/d \right).
\end{equation}
Taking into account (\ref{6}), (\ref{30}), (\ref{31}),
we get a condition under which one can neglect this term:
\begin{equation}
\label{7}
\frac{d^3}{\alpha_p}
\gg
\frac
{
1.71 C
\ln
\left( 1+L/d \right)
}
{
\left[
    1-1.71 C \ln(L/\sigma)
\right]
}
.
\end{equation}
Note that the condition (\ref{7})
does not depend on the way of excitation of the surface atoms.
It remains the same if the atoms are excited by a wave incident
from the side of the probe, by an evanescent wave, or by the wave
radiated by the probe itself.

In eq.(\ref{7}) $\ln(L/d+1)$ and $\ln(L/\sigma)$ are of the order of
unity. Therefore, one can rewrite (\ref{7}) in a simpler form
\begin{equation}
\label{est-1}
\frac{d^3}{\alpha_p}
\gg
\frac
{1.71 C}
{1-1.71 C}
.
\end{equation}
Usually for dielectric media $C\sim 0.1$ and $d^3/\alpha_p \sim 10$
at the distances from the surface equal to one lattice constant.
Therefore, inequality (\ref{est-1}) is satisfied at the distances
equal or greater than one lattice constant.

At the distances between the probe and the surface such that the influence
of discrete atomic distribution is negligible but the retardation
is still important the calculation of the field
on the probe reduces to the electrostatic problem of the
calculation of the dipole field near
a dielectric surface. For the field ${\bf E}_R^{eff}$ at the probe position
we get
\begin{equation}
\label{31-far}
{\bf E}_R^{eff}
\approx
\frac{\pi}4 \frac{C}{1+\frac23 \pi C}
\frac{\alpha_p}{d^3}\left({\bf E}_p-3({\bf E}_p {\bf n}){\bf n}\right),
\end{equation}
where ${\bf n}=(0,0,1)$ is a unit vector perpendicular to the monolayer.
We write here ``$\approx$'', because in the derivation of eq.(\ref{31-far})
we have used the Lorentz-Lorenz formula for the refractive index
$n$:
\begin{equation}
\label{LL}
n^2=\frac{1+(8\pi /3)C}{1-(4\pi /3)C}
\quad,
\end{equation}
which does not describe properly optical properties in
the surface region~\cite{GKJOSA}. The condition which allows to
neglect the last term in eq.(\ref{E-p}) is given now by
\begin{equation}
\label{est-2}
\frac{d^3}{\alpha_p}
\gg
\frac{\pi}4
\frac{C}{1+\frac23 \pi C}
.
\end{equation}
The condition (\ref{est-2}) is in a good agreement with (\ref{est-1}),
albeit the former imposes less strict limitations on $d$.

Thus, at the distances between the probe and the surface of the order
of one lattice constant the back influence of the film on the probe
can be safely neglected. In this approximation the field on the probe
is determined by the field of the external wave and the field
${\bf E}_l^0$.
The problem of calculation of the field ${\bf E}_l^0$ will be discussed
in the next section.

\section
{
Distribution of the field inside the film
(solution of the unperturbed problem)
}

Let the external wave be a plane one, i.e.,
\begin{equation}
\label{plain}
{\bf A}_I({\bf r})
=
{\bf E}_{0I}
\exp( i {\bf k}_0 {\bf r} )
,\;
{\bf k}_0 = k_0
(\sin\Theta_I\cos\Phi_I,\sin\Theta_I\sin\Phi_I,-\cos\Theta_I)
\;,
\end{equation}
where $\Theta_I$ is an incident angle and the azimuthal angle
$\Phi_I$ defines the orientation of the incidence plane.

The atoms of the monolayers are assumed to form a regular periodic
structure with elementary translations vectors
${\bf a}_1$ and ${\bf a}_2$. Without loss of generality one can put
\begin{equation}
{\bf a}_1 = a_\|(1,0,0)
,\quad
{\bf a}_2 = a_\|(\alpha,\beta,0)
.
\end{equation}
For the sake of simplicity the vectors ${\bf a}_1$ and ${\bf a}_2$
are assumed to be the same for all monolayers. The generalization
on different ${\bf a}_1$ and ${\bf a}_2$ for different monolayers
can be easily done.
In these notations the radius-vector of some atom in the
$j$-th monolayer has the form
\begin{equation}
\label{raj}
{\bf r}_{a_j}
=
{\bf r}_j^0 + {\bf a}_{mn}
=
{\bf r}_j^0 + m {\bf a}_1 + n {\bf a}_2
\;,
\end{equation}
where $m$ and $n$ are integer. The vector ${\bf r}_j^0$ has the following
form
\begin{equation}
{\bf r}_j^0
=
\alpha_{1j} {\bf a}_1 + \alpha_{2j} {\bf a}_2 + z_j {\bf n}
\;,\quad
0 \le \alpha_{1(2)j} < 1
\;.
\end{equation}
Nonvanishing $\alpha_{1j}$ and $\alpha_{2j}$ take into account possible
parallel shifts of the atomic planes relative to one another.

Making use of the principle of parallel translational symmetry
the solution for the local field in the
$l$-th monolayer
can be written down in the form~\cite{PWS,PW,WSM,WE,Wijers}:
\begin{equation}
\label{sol-m}
{\bf E}^0_l({\bf r}^\|)
=
{\bf E}^0_l
\exp
\left(
     i {\bf k}_0 {\bf r}^\|
\right)
.
\end{equation}

We substitute (\ref{plain}), (\ref{sol-m}) into (\ref{E-0}).
Now it is necessary to calculate the lattice sums in eq.(\ref{E-0}).
Due to the slow convergence of lattice sums their direct calculation
is a very laborious and in fact impracticable computational problem.
In order to compute them one can use ``the Lorentz method" according
to which the atoms located near the observation point are treated
as discretely distributed and the atoms outside a fictitious boundary
are treated as continuously distributed. This method can be employed
either for well-ordered or for random media and for the observation
points either inside or outside of the media. This way of computation
has been employed in various modifications in
Refs.~\cite{KSJPB,KS, WSM,GKJOSA,Gadom}.
On the other hand, if the atomic distribution in the monolayers
is regular, this separation of discrete and continuous regions
is not necessary anymore. In this case there exist another way
of computation
of the lattice sums which is rapidly convergent and more convenient
for the interpretation of the field behavior near
the surface~\cite{PWS,Wijers}.
It is this latter way will be employed in the present paper.

If the observation point is located not in the monolayer
over which the summation is carried out ($j \ne l$)
the lattice sums can be converted to the form which is more
appropriate for numerical calculations, making use of the formula
for the dipole field induced by the $j$-th monolayer
at some point ${\bf r}$, which has been obtained in Ref.~\cite{Wijers}
by means of Fourier transform
(Ewald's three-fold integral transform):
\begin{eqnarray}
\label{F}
{\bf E} ({\bf r},z_j)
&=&
\alpha \sum_{a_j}
\nabla_{\bf r}\times\nabla_{\bf r}\times
{\bf E}_j^0
\exp
\left(
    i{\bf k}_0 {\bf r}_{a_j}^\|
\right)
G
\left(
    R_{a_j}
\right)
\nonumber\\
&=&
\sum_{p,q=-\infty}^{\infty}
{\bf A}_{pq}({\bf r},z_j)
\exp
(
    i{\bf k}_0^\| {\bf r}
)
\;,\;
z \ne z_j
\;,
\\
{\bf A}_{pq}({\bf r},z_j)
&=&
-
\frac
{2\pi i\alpha}
{
  \left|
      {\bf a}_1
      \times
      {\bf a}_2
  \right|
}
\left[
   {\bf k}_{pq}
   \times
   \left(
        {\bf k}_{pq}
        \times
        {\bf E}^0_j
   \right)
\right]
\frac
{
\exp
\left[
    i ({\bf k}_{pq}-{\bf k}_0^\|)
    ({\bf r}-{\bf r}^0_j)
\right]
}
{\kappa_{pq}}
\;,
\nonumber
\end{eqnarray}
\begin{eqnarray}
{\bf k}_{pq}
&=&
\left\{
\begin{array}{lcl}
({\bf k}_0^\|+{\bf g}^{\|}_{pq}, \kappa_{pq})
&{\rm if}&
z > z_j
\\
({\bf k}_0^\|+{\bf g}^{\|}_{pq},-\kappa_{pq})
&{\rm if}&
z < z_j;
\end{array}
\right.
\\
\label{kappa}
\kappa_{pq}
&=&
\sqrt{k_0^2-({\bf k}_0^\|+{\bf g}^{\|}_{pq})^2}
,\quad
{\bf g}^{\|}_{pq}=p{\bf g}_1+q{\bf g}_2
\\
{\bf k}_0^\|
&=&
k_0
( \sin\Theta_I\cos\Phi_I , \sin\Theta_I\sin\Phi_I , 0 )
\nonumber
.
\end{eqnarray}
The vectors of the reciprocal lattice ${\bf g}_1$ and ${\bf g}_2$
are related to the vectors ${\bf a}_1$ and ${\bf a}_2$ as
\begin{equation}
{\bf g}_1
=
2\pi \frac{{\bf a}_2\times {\bf n}}{|{\bf a}_1\times {\bf a}_2|}
,
\qquad
{\bf g}_2
=
2\pi \frac{{\bf n} \times {\bf a}_1}{|{\bf a}_1\times {\bf a}_2|}
.
\end{equation}
The choice of the vectors ${\bf a}_1$ and ${\bf a}_2$ can be arbitrary.
However, it makes sense to choose them in such a manner that they
have a minimal length. Then the length of the vectors
${\bf g}_1$ and ${\bf g}_2$ will be also minimal. Such a choice of
the basis is convenient, because in this case the decay coefficients
of the evanescent waves for square lattices and for slightly anisotropic
lattices increase with the parameter $\kappa=p^2+q^2$.

Formula (\ref{F}) is a useful representation of the lattice sums
for numerical calculations because the sums over $p$, $q$
converge much faster than the sums over $a_j$, provided that
$|z-z_j| \ge a_\|$. The calculations carried out in the next section
show that in the sum over $p$, $q$ it is enough to keep only few
summands with minimal ${g}_{pq}^{\|}$.

On the other hand, formula (\ref{F})  has a very clear physical
interpretation. It represents itself a decomposition of the field
induced by the monolayer of discrete atoms on propagating and evanescent
waves. Indeed, if
$
\left|
   {\bf k}_0^\| + {\bf g}_{pq}^\|
\right|
<
k_0
$
the quantity $\kappa_{pq}$ is real and we have propagating waves. Otherwise,
$\kappa_{pq}$ is imaginary and we have exponentially decaying waves.
A more detailed discussion of the role of propagating and evanescent waves
will be given in the next section.

Note that the quantities ${\bf A}_{pq}({\bf r},z_j)$ with
$p,q \ne 0$, i.e. the evanescent harmonics, are periodic functions
of $x$ and $y$ and they posses the following property of the translational
symmetry
\begin{equation}
{\bf A}_{pq}({\bf r} + m {\bf a}_1 + n {\bf a}_2,z_j)
=
{\bf A}_{pq}({\bf r},z_j)
\;.
\end{equation}
As we shall see further on, this property of the evanescent harmonics
leads to the fact that the intensities of the reflected and transmitted
waves near the film surface are periodic functions of $x$ and $y$.

If the observation point ${\bf r}$ is at the position of some atom inside the
monolayer over which the summation is carried out ($j=l$),
then for the calculation of the lattice sum it is convenient to
employ the method based on Ewald's onefold integral transform~\cite{PWS}.
Separating the initial sum over the real lattice on the sums over the
real and reciprocal lattices one gets the following expression for the
dipole field~\cite{PWS}
\begin{equation}
\label{d+c}
\alpha
\sum_{a_j}
\nabla_{\bf r}
\times
\nabla_{\bf r}
\times
\left.
{\bf E}_j^0
\exp
\left(
    i{\bf k}_0 {\bf r}_{a_j}^\|
\right)
G
\left(
    R_{a_j}
\right)
\right|_{z=z_j}
+
i \frac{2}{3} k_0^3 \alpha
{\bf E}_j^0
\exp
(
      i {\bf k}_0^\| {\bf r}
)
=
\alpha
\hat f({\bf k}_0)
{\bf E}_j^0
\exp
(
      i {\bf k}_0^\| {\bf r}
)
\;,
\end{equation}
where the term with ${\bf r}^\| = {\bf r}_{a_j}^\|$ has to be excluded
from the summation.
Tensor $\hat f$ is a symmetric one and the components
$f_{xz}$, $f_{yz}$, $f_{zx}$, $f_{zy}$ vanish. It has the form~\cite{PWS}
\begin{eqnarray}
\label{tensor-f}
f^{\nu\mu}({\bf k})
&=&
    c^{\nu\mu}
    -
    \left[
        \frac{2}{3}
        i k^3
        {\rm erfc}
        \left(
            \frac{i k}{2 E}
        \right)
        +
        \frac{4 E}{3 \sqrt{\pi}}
        \left(
            k^2 - E^2
        \right)
        \exp
        \left(
            \frac{k^2}{4E^2}
        \right)
    -
    \frac{2}{3} i k^3
    \right]
    \delta_{\nu\mu}
\nonumber\\
c^{\nu\mu}
&=&
\frac
{i \pi}
{
  \left|
      {\bf a}_1 \times {\bf a}_2
  \right|
}
\sum_{p,q}
\left[
    \frac
    {
      k^2 \delta_{\nu\mu}
      -
      k_{pq\nu} k_{pq\mu}
    }
    {k_{pq}}
    \Delta_{pq}
    ( 1 + \tau )
    +
    \eta \Sigma_{pq}
\right]
\nonumber\\
&&
+
\frac{1}{2}
\sum_{n,m}
\frac
{
  \exp( i {\bf k} {\bf a}_{nm} )
}
{a_{nm}^3}
\left\{
     \Gamma_{nm}^1
     \left[
         \delta_{\nu\mu}
         \Gamma_{nm}^2
         +
             {\bf a}_{nm}^\nu
             {\bf a}_{nm}^\mu
         \Gamma_{nm}^3
     \right]
\right.
\nonumber\\
&&
\left.
     +
     \Gamma_{nm}^4
     \left[
         - \delta_{\nu\mu} a_{nm}
         +
             {\bf a}_{nm}^\nu
             {\bf a}_{nm}^\mu
         \Gamma_{nm}^5
     \right]
     +{\rm c.c.}
\right\}
,
\end{eqnarray}
where
$\nu,\mu=x,y,z$, $a_{nm}=|{\bf a}_{nm}|$
\begin{displaymath}
\eta=\delta_{\mu z}\delta_{\nu z},\quad
\tau=(-1)^{\delta_{\mu z}} (-1)^{\delta_{\nu z}}
\end{displaymath}
\begin{displaymath}
\Delta_{pq}
=
{\rm erfc}
\left(
    -
    \frac{i k_{pq}}{2 E}
\right)
\quad,\quad
\Sigma_{pq}
=
\frac{i4E}{\sqrt{\pi}}
\exp
\left(
    \frac{k_{pq}^2}{4E^2}
\right)
\end{displaymath}
\begin{displaymath}
\Gamma_{nm}^1
=
\exp
\left(
    - i k a_{nm}
\right)
{\rm erfc}
\left(
     a_{nm} E
     -
     \frac{ik}{2E}
\right)
\quad,\quad
\Gamma_{nm}^2
=
- 1 - i k a_{nm} + k^2 a_{nm}^2
\quad,
\end{displaymath}
\begin{displaymath}
\Gamma_{nm}^3
=
- k^2 + \frac{3 i k}{a_{nm}} + \frac{3}{a_{nm}^2}
\quad,\quad
\Gamma_{nm}^4
=
\frac{2E}{\sqrt{\pi}}
\exp
\left(
    - E^2 a_{nm}^2 + \frac{k^2}{4E^2}
\right)
\quad,\quad
\Gamma_{nm}^5
=
\frac{3}{a_{nm}} + 2 E^2 a_{nm}
.
\end{displaymath}
The rate of convergence of the sums in (\ref{tensor-f}) is
determined by the parameter $E$, which has a dimension of the inverse
length. It appears as a result of the formal separation of the initial sum
over the real lattice on the sums over the real and reciprocal lattices.
This parameter has no any physical meaning, it must be only real and positive
and its choice can be arbitrary~\cite{PWS}. However,
it makes sense to choose the parameter $E$ in such a manner
that the rates of convergence for the sums over $(p,q)$ and $(m,n)$
are the same. One can show that this
requirement is fulfilled, provided that
$E=\sqrt{\pi /|{\bf a}_1\times {\bf a}_2|}$.
Substituting this value into (\ref{tensor-f})
one can estimate the maximal values of $(n,m)$ and $(p,q)$
which has to be taken for the calculation of the lattice sums
with the tolerance $\varepsilon$.
Taking into account that at large $x$
${\rm erfc}(x)\approx \exp(-x^2)/(x\sqrt{\pi})$, we get the condition
\begin{equation}
\label{max-mn}
\exp(-\pi a_{nm}^2/|{\bf a}_1\times {\bf a}_2|)
\approx
\varepsilon
.
\end{equation}
Exact numerical calculations for a square lattice show that if
the summation is restricted by the terms with
$n,m,p,q=0,\pm 1$ the relative tolerance of the calculations
of the dipole field is about $10^{-6}$, which is in a good agreement
with the estimate (\ref{max-mn}). Therefore, our choice of the parameter
$E$ appears to be very convenient for carring out numerical calculations.

Note that
in the computer calculations of the dipole field the algorithm
based on Ewald's onefold integral transform is 5-10 times faster
than that based on the Lorentz method.

Taking into account identities (\ref{F}), (\ref{d+c})
the system of equations (\ref{E-0}) can be rewritten in the following form:
\begin{eqnarray}
\label{field-l}
\left[
    1 - \alpha \hat f({\bf k}_0)
\right]
{\bf E}_l^0
&=&
{\bf E}_{Il}
-
i2\pi C a
\sum_{j=1}^{N}{}'
\sum_{p,q=-\infty}^{\infty}
\left[
    {\bf k}_{pq}
    \times
    \left(
        {\bf k}_{pq}
        \times
        {\bf E}_j^0
    \right)
\right]
\frac{\exp
\left[
     i
     \left(
         {\bf k}_{pq}
         -
         {\bf k}_0^\|
     \right)
     \left(
          {\bf r}_l^0-{\bf r}_j^0
     \right)
\right]}{\kappa_{pq}}
\;,
\nonumber\\
&&l=\overline{1,N}
\;,
\end{eqnarray}
where $C=\alpha/(a|{\bf a}_1 \times {\bf a}_2|)$,
\begin{equation}
{\bf E}_{Il}
=
{\bf E}_{0I}
\exp
\left(
     i {\bf k}_0^\perp {\bf r}_{a_l}
\right)
=
{\bf E}_{0I}
\exp
\left(
     - i k_0 \cos\Theta_I z_l
\right)
,\quad
l=\overline{1,N}
.
\end{equation}
Prime at the sum sign in (\ref{field-l}) means that the summand with
$j=l$ is excluded.

\begin{figure}
\begin{center}
\leavevmode
\begin{tabular}{cc}
\epsfxsize=8cm \epsffile{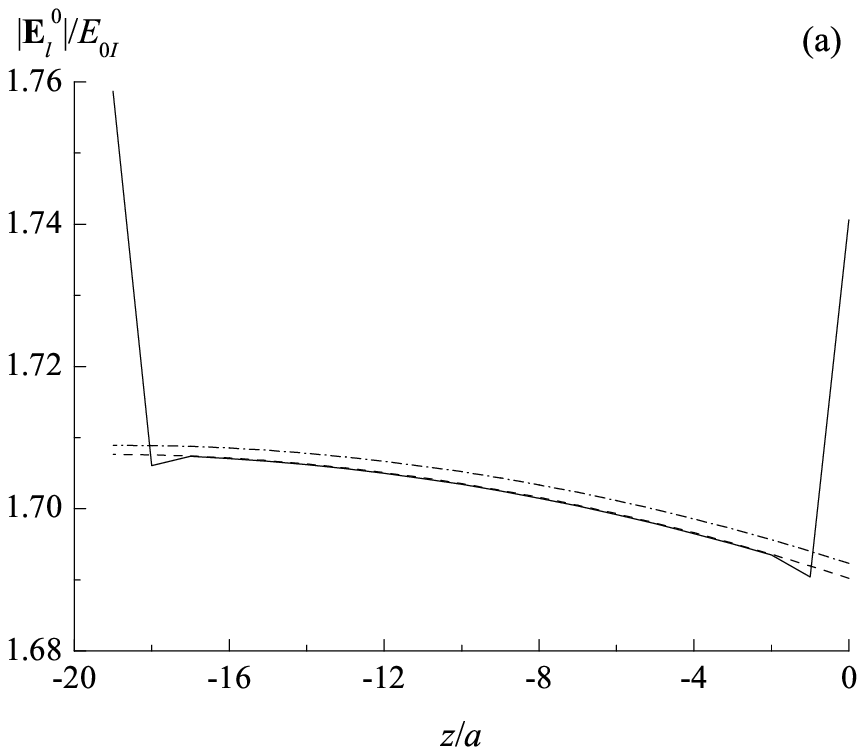} & \epsfxsize=8cm \epsffile{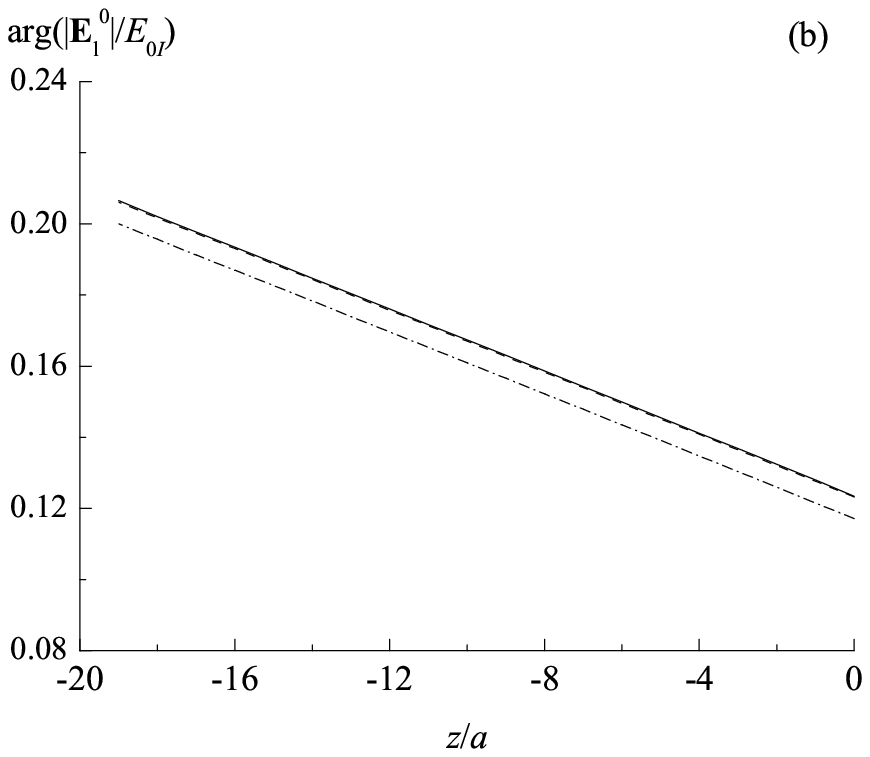} \\
\epsfxsize=8cm \epsffile{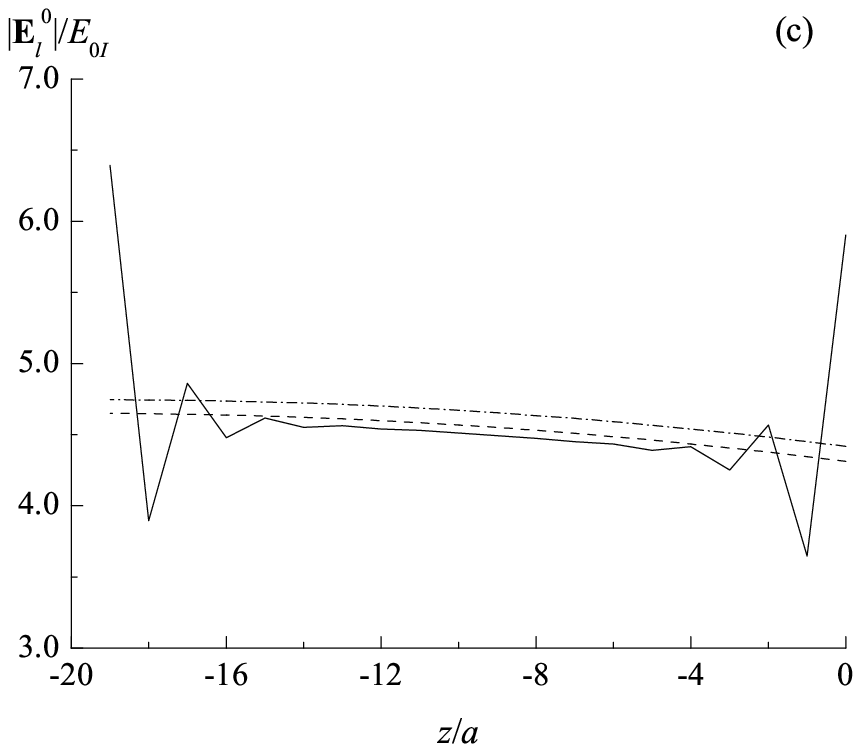} & \epsfxsize=8cm \epsffile{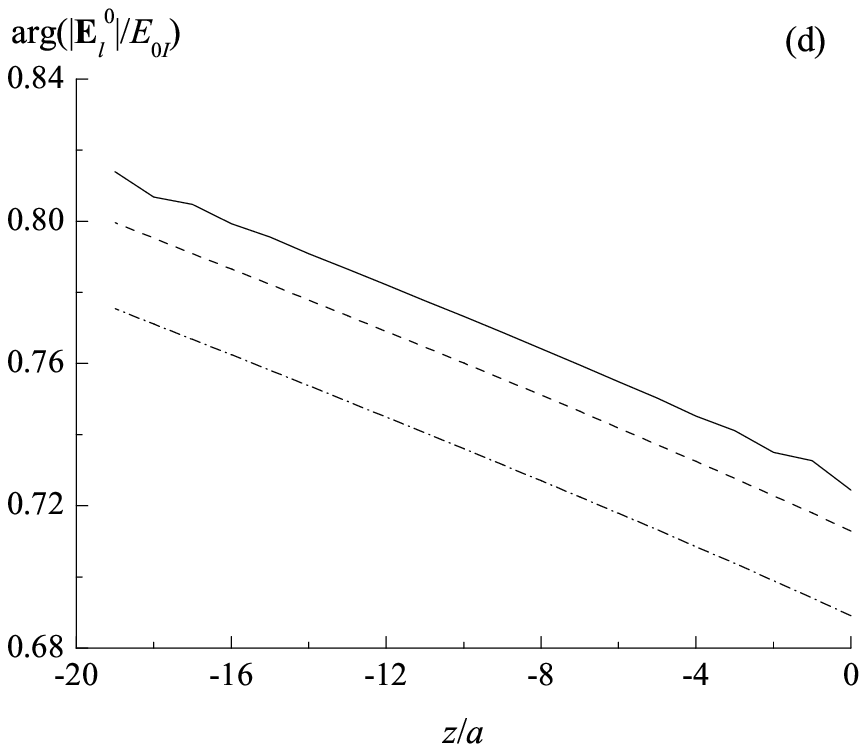}
\end{tabular}
\end{center}
\caption
{
Coordinate dependences of the amplitude (a, c) and the phase
(b, d) of the local electric field strength inside the film:
------ numerical solution of the system of equations (33),
$\cdot\ -\ \cdot\ - $ calculation according to the Airy formulae for
the case when the film occupies the region of space $-a(N-1)<z<0$,
$-\ -\ -$ calculation according to the Airy formulae for
the case when the film occupies the region of space $-a(N-1/2)<z<a/2$.
The calculations are performed for the cubic lattice. The external
wave is polarized along one of the translation vector of the lattice
($s$-polarization). The parameters are:
$k_0 = 0.01\ {\rm nm}^{-1}$, $a = 0.5\ {\rm nm}$,
$N=20$, $\Theta_I=30^o$, $C=0.1$ (a, b); $C=0.2$ (c, d).
}
\label{fig_inside}
\end{figure}

The results of numerical solution of the system of equations (\ref{field-l})
for the cubic lattice and $s$-polarized incident wave are shown in
Fig.~\ref{fig_inside}.
In order to demonstrate to the fullest extent the characteristic features
of the field behavior near the film surface we have used a rather high
value of the volumic polarizability ($C=0.2$). Normally in the optical range
$C = 0.04 \div 0.15$ which corresponds to the refractive index
$n = 1.27 \div 2.46$~\cite{Yariv}. However, for some materials $C$
can be higher. For instance,
$n_{GaP}(\lambda_0=564\ nm)=3.42$~\cite{PWS-SSC,AS} and
$n_{Si}(\lambda_0=620\ nm)=3.94$~\cite{WE,AS} which corresponds to
$C_{GaP}=0.19$ and $C_{Si}=0.2$, respectively.

In Fig.~\ref{fig_inside} the results of the calculations according to
the Airy formulae are also shown for comparison.
Besides we have used the Airy formulae for two cases:
(1) the film occupies the region of space $-a(N-1)<z<0$, and
(2) the film occupies the region of space $-a(N-1/2)<z<a/2$.
In the former case the corresponding formula has the
form~\cite{BW}
\begin{equation}
\label{Eya}
E^y({\bf r})
=
\frac
{
 t_\perp
 \left[
       \exp
       \left(
             i k_0n {\bf s}_T^{(-)} {\bf r}
       \right)
       -r_\perp
       \exp (i\varphi)
       \exp
       \left(
             i k_0n {\bf s}_T^{(+)} {\bf r}
       \right)
 \right]
}
{
1-r_\perp^2
\exp(i\varphi)
}
\frac{E_{0I}^y}{1-(4\pi /3)C}
,
\end{equation}
\begin{equation}
\label{st}
{\bf s}_T^{(\pm )}=
\left(
      -\sin \Theta_T,0, \pm \cos \Theta _T
\right)
,
\varphi = 2 k_0 h n \cos \Theta_T
,
\end{equation}
where $\Theta_T$ is the angle of refraction, $t_\perp $ and $r_\perp$ are
Fresnel coefficients of transmittance and reflectance for s-polarization,
respectively, $h=a(N-1)$
is a film thickness, $n$ is a refractive index
defined by the Lorentz-Lorenz relation (\ref{LL}).

In the latter case $h=aN$, and in eq.(\ref{Eya}) one has to make
the replacement ${\bf r} \to {\bf r} - {\bf n} a/2$,
$E_{0I}^y \to E_{0I}^y \exp(- i k_0 a \cos\Theta_I/2)$.
It seems that the first form of the Airy formulae better
corresponds  to the initial statement of the problem.
However, the results of
numerical solution of the system (\ref{field-l}) are in a better agreement
with the second form. The reason is that in the former case
the macroscopic density $\rho$ which enters the Airy formulae appears
to be higher than its true value
$
\left(
    a|{\bf a}_1 \times {\bf a}_2|
\right)^{-1}
$,
but in the latter case $\rho$ in the Airy formulae is equal to
$
\left(
    a|{\bf a}_1 \times {\bf a}_2|
\right)^{-1}
$
as it should be.

As it follows from the numerical calculations, the consideration of
the discrete structure influences the most significantly the behavior of
the field amplitude near the film surface where rapid oscillations
of the local field take place. The amplitude of these oscillations
and the penetration depth into the film volume increase with the
parameter $C$. One can also see on Fig.~\ref{fig_inside}
that the consideration of the discrete structure does not influence much
the phase of the field. This means that the wave vector of the field
remains almost the same at any point inside the film. Therefore, the real
part of the refractive index of the field is constant. Far from the
boundaries the distribution of the field is regular and approximately
coincides the results given by the Airy formulae.

\section{Field on the probe}

When the system of equations (\ref{field-l}) is solved and all the
quantities ${\bf E}_j^0$, $j=\overline{1,N}$ are determined, one can
calculate the field at the probe position:
\begin{equation}
\label{32}
{\bf E}({\bf r})
=
{\bf A}_I({\bf r}) -
2\pi i C a
\sum_{j=1}^N
\sum_{p,q=-\infty}^{\infty}
\left[
    {\bf k}_{pq}
    \times
    \left(
        {\bf k}_{pq}
        \times
        {\bf E}^0_j
    \right)
\right]
\frac
{
  \exp
  \left[
      i {\bf k}_{pq}
      \left({\bf r}-{\bf r}^0_j\right)
  \right]
}
{\kappa_{pq}}
  \exp
  \left(
      i
      {\bf k}_0^\|
      {\bf r}^0_j
  \right)
.
\end{equation}
Under the condition (\ref{7}) the field ${\bf E}({\bf r})$
equals to the field ${\bf E}_p$ acting on the probe.
Let's analyze in details the expression (\ref{32}).

\subsection{Field in the wave zone}

Let's consider first the case when the observation point is in the
wave zone, i.e., $|z-z_j| \to \infty$. In this case the summands with
$p,q \ne 0$ do not contribute to the resultant field due to the following
reason. All the quantities ${g}_{pq}^{\|}$ at $p,q\not= 0$ are
of the order of $2\pi/a$ and they are much greater than $k_0$
if $k_0 a \ll 1$. Therefore, $\kappa_{pq}$ (\ref{kappa}) are purely
imaginary and all the corresponding summands in (\ref{F}) decay rapidly
with the increase of the distance from the surface. Only the term with
${\bf k}_{00}$ survives. If $z>z_j$, ${\bf k}_{00}$ represents itself
the wave vector of the reflected wave ${\bf k}_R$, otherwise
${\bf k}_{00}={\bf k}_0$. Therefore, in the wave zone we have plane
propagating waves:
\begin{eqnarray}
{\bf E}({\bf r})
&=&
\left\{
\begin{array}{rcl}
{\bf A}_I({\bf r})
+
{\bf E}_R^{far}
\exp
\left(
     i {\bf k}_R {\bf r}
\right)
&,&
z > 0;
\\
{\bf E}_T^{far}
\exp
\left(
     i {\bf k}_0 {\bf r}
\right)
&,&
z < - a(N-1);
\end{array}
\right.
\\
{\bf E}_R^{far}
&=&
- i 2 \pi C
\frac{a}{k_0\cos\Theta_I}
\sum_{j=1}^{N}
\left[
     {\bf k}_R \times
     \left(
          {\bf k}_R \times {\bf E}_j^0
     \right)
\right]
\exp
\left(
    - i {\bf k}_R^\perp {\bf r}^0_j
\right)
\;,
\nonumber\\
{\bf E}_T^{far}
&=&
{\bf E}_{0I}
- i 2 \pi C
\frac{a}{k_0\cos\Theta_I}
\sum_{j=1}^{N}
\left[
     {\bf k}_0 \times
     \left(
          {\bf k}_0 \times {\bf E}_j^0
     \right)
\right]
\exp
\left(
     - i {\bf k}_0^\perp {\bf r}^0_j
\right)
\;.
\nonumber
\end{eqnarray}
This fields are transverse, because
${\bf k}_R {\bf E}_R^{far} = {\bf k}_0 {\bf E}_T^{far} = 0$.
The amplitudes
${\bf E}_R^{far}$, ${\bf E}_T^{far}$ do not depend on the coordinates.

\begin{figure}
\epsfxsize=8cm \centerline{\epsffile{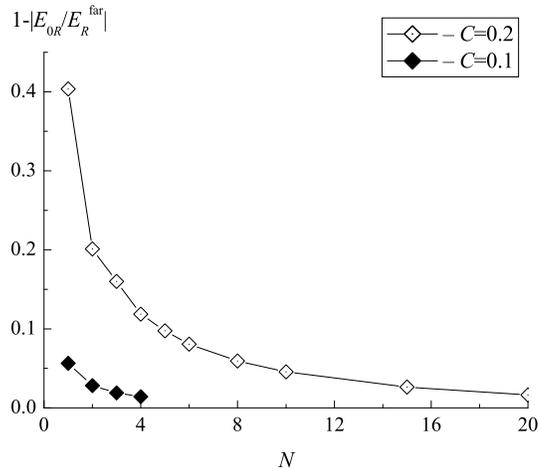}}
\caption
{
The difference between the reflected wave field in the near zone,
calculated under ($E_R^{far}$) and without ($E_{0R}$) the consideration
of the discrete
structure of the medium, as a function of the number of monolayers $N$
which constitute the film.
The calculations are performed for the cubic lattice.  The parameters are:
$k_0 = 0.01\ {\rm nm}^{-1}$, $a = 0.5\ {\rm nm}$, $\Theta_I=0^o$.
}
\label{fig_RTN}
\end{figure}

\begin{figure}
\epsfxsize=8cm \centerline{\epsffile{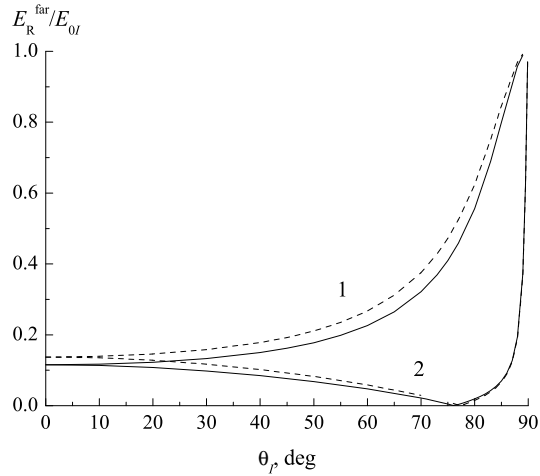}}
\caption
{
The dependence of the electric field strength amplitude of the reflected
wave $E_R^{far}$ on the incident angle of the external wave
$\Theta_I$ in the wave zone for $s$- (1) and p- (2) polarizations:
------ under consideration of the discrete structure,
-- -- -- according to the Airy formulae.
The calculations are performed for the cubic lattice.  The parameters are:
$k_0=0.01\ {\rm nm}^{-1}$, $a=0.5\ {\rm nm}$, $N=3$, $C=0.2$.
}
\label{fig_RT}
\end{figure}

The discrepancy between the values of the field of the reflected
wave calculated under consideration of the discrete structure and
according to the Airy formulae can be of the order of $10\%$.
This discrepancy becomes larger with the decrease of the film thickness
and with the increase of the parameter $C$ (Fig.~\ref{fig_RTN}).
The plots on Fig.~\ref{fig_RTN} lead to the conclusion that at high
enough film thicknesses ($N \ge 20$) the Airy formulae give a
satisfactory description of the reflected waves in the wave zone.
In this situation our method of calculation becomes equivalent to
that developed in Refs.~\cite{JS89,JSS,JS90}.
As for the dependence on the incident angle, the maximal discrepancy
between the amplitudes of the reflected wave calculated under consideration
and without consideration of the discrete structure is observed at
normal incidence. With the increase of the incident angle this
discrepancy gradually vanishes (Fig.~\ref{fig_RT}).
In the calculations presented on Figs.~\ref{fig_RTN},\ref{fig_RT}
we used the Airy formulae for the case when the film occupies the region
of the space $-a(N-1/2) < z < a/2$~\cite{BW}
\begin{equation}
E_{0R}^y
=
\frac
{
  r_\perp
  \left[
        1-\exp(i\varphi)
  \right]
}
{
 1 - r_\perp^2 \exp(i\varphi)
}
\exp
\left(
     -i k_0 a \cos\Theta_I
\right)
E_{0I}^y
.
\end{equation}
Here we use the same notations as in formula (\ref{Eya}).
Besides, as for the field inside the film, the results of our calculations
are in a better agreement with the Airy formulae compared to the case when
the film occupies the region of space $-a(N-1) < z < 0$.
We would like to note also that for $p$-polarized wave the field
of the reflected wave near the Brewster angle does not vanish~\cite{GS2000}.

\subsection{Field in the near zone}

If the observation point is near the surface, one can observe a number
of characteristic features in the field behavior which we are going to
discuss now. For the sake of simplicity we shall restrict ourselves, if the
opposite is not stated explicitly, by the case of quadratic lattice
(${\bf a}_1 \perp {\bf a}_2$, $a_1=a_2=a$; $g_1=g_2=g=2\pi/a$).

\subsubsection
{
Contributions of propagating and evanescent harmonics
}

\begin{figure}
\epsfxsize=8cm \centerline{\epsffile{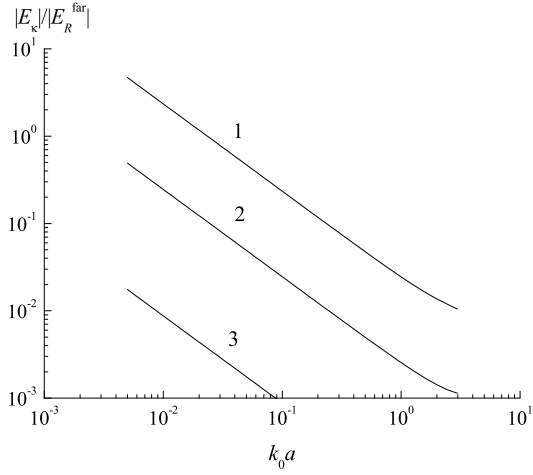}}
\caption
{
The ratio of the evanescent field amplitudes near the surface $E_{\kappa}$
to the amplitude of the reflected field in the wave zone $E_R^{far}$.
The calculations are performed for the cubic lattice.  The parameters are:
$k_0=0.01 {\rm nm}^{-1}$, $\Theta_I=0$, the distance between the
observation point and the surface is equal to $a$.
1 -- $|E_1/E_R^{far}|$, 2 -- $|E_2/E_R^{far}|$, 3 -- $|E_4/E_R^{far}|$.
}
\label{fig1}
\end{figure}

\begin{figure}
\epsfxsize=8cm \centerline{\epsffile{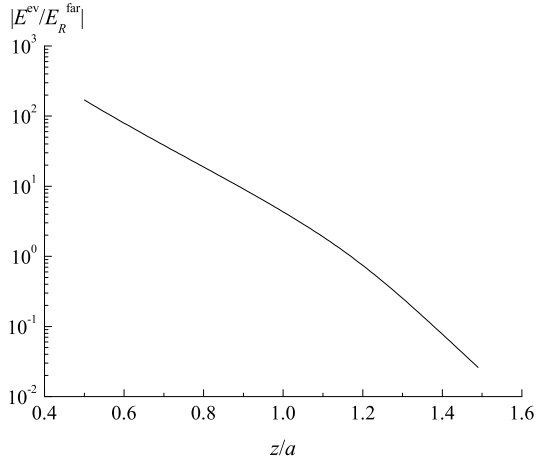}}
\caption
{
Coordinate dependences of the ratio of the total amplitude of the
evanescent waves $E^{ev}$ to the amplitude of the reflected wave
field in the wave zone $E_R^{far}$.
The calculations are performed for the cubic lattice.  The parameters are:
$k_0 = 0.01\ {\rm nm}^{-1}$, $a = 0.5\ {\rm nm}$, $\Theta_I=0^o$.
The observation point is above an atom of the film.
}
\label{fig_near}
\end{figure}

At a small distance from the surface the summands with
$p,q \ne 0$ in (\ref{32}) can make a noticeable contribution. Let's
estimate this contribution considering the following example. As it was
mentioned above in the discussion of formula (\ref{F}), it is
convenient to single out in eq.(\ref{32}) evanescent waves
$
  {\bf E}_\kappa
  =
  \sum_{p^2+q^2=\kappa}
  {\bf A}_{pq}
  \exp
  \left(
      i
      {\bf k}_0^\|
      {\bf r}
  \right)
$
with the decay coefficients dependent on $\kappa$.
The ratio of the amplitudes of the evanescent harmonics
$E_{\kappa}$ for $\kappa=0$ (plane reflected wave), $1$, $2$, $4$ is
shown on Fig.~\ref{fig1}.
Note that the relative contribution of different $E_{\kappa}$
strongly depends on the distance between the probe and the film surface.
The behavior shown on Fig.~\ref{fig1} is typical for the distances
greater or equal to one lattice constant. Usually if one takes into
account the modes with $p,q=0,\pm 1,\pm 2$
the tolerance of the calculations is less than one per cent.
As it follows from the dependences on Fig.~\ref{fig1}, at
$k_0 a=\pi$ (a typical value for photonic crystals and optical
lattices) the monolayers do not fill the evanescent waves radiated
by the other monolayers. At $z=2a$ and $k_0 a=0.005$ the ratio of decaying
and non-decaying field components is about $1\%$ and rapidly decreases
with the increase of $k_0 a$. If $a \ll 1/k_0$, then $g \gg k_0$,
and as it follows from eqs.(\ref{F})-(\ref{kappa})
the form of the evanescent waves is almost independent of the incidence
angle of the external wave. Coordinate dependences of the total amplitude
of the evanescent harmonics is shown on Fig.~\ref{fig_near}.

\begin{figure}
\epsfxsize=8cm \centerline{\epsffile{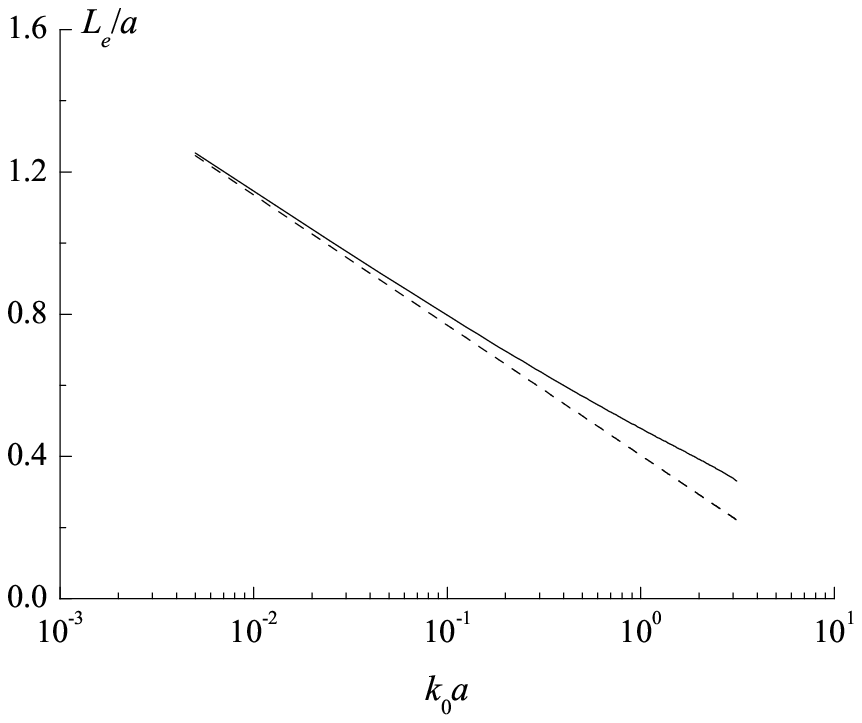}}
\caption{
The dependence of the distance $L_e$, at which the amplitudes of the
evanescent and propagating parts of the reflected wave field are equal
to one another, on the parameter $k_0 a$:
\mbox{------} exact numerical calculations,
$-\ -\ -$ an estimate according to formula (42).
The calculations are performed for the cubic lattice, $\Theta_I=0^o$.
}\label{fig2}
\end{figure}

The dependence of the distance $L_e$, at which the amplitudes of the
evanescent and propagating components are equal, on the parameter
$k_0 a$ is depicted on Fig.~\ref{fig2}. 
Due to the rapid decay of the
evanescent harmonics the quantity $L_e$ can be treated as a distance
from the surface up to which the contribution of the evanescent component
to the resultant field is essential. From the dependence shown on
Fig.~\ref{fig2} it follows that the contribution of the evanescent
component is essential at the distances less than two lattice constants.
The field radiated by the monolayers, which are located at larger distances
from the observation point, can be treated as propagating plane waves.

The distance $L_e$ can be also analytically estimated on the basis
of the following consideration. As it follows from the dependences
depicted on Fig.~\ref{fig1} the leading contribution to the decaying
wave is made by the evanescent wave with $\kappa=1$. Therefore, the
distance $L_e$ can be approximately determined from the condition
\begin{equation}
\label{condition}
\frac
{\left| E_1 \right|}
{\left| E_R^{far} \right|}
=1.
\end{equation}
In the case $k_0 a \ll 1$ we have
${\bf k}_{pq} \approx ({\bf g}_{pq}^\|,ig_{pq}^\|)$.
Then for a monolayer we can write the condition (\ref{condition})
anew as
\begin{equation}
2
\frac{2\pi}{k_0 a}
\exp
\left(
    - 2\pi \frac{L_e}{a}
\right)
=1
,
\end{equation}
from which we get
\begin{equation}
\label{L_e-a}
\frac{L_e}{a}
=
\frac{1}{2\pi}
\ln
\left(
    \frac{4\pi}{k_0 a}
\right)
.
\end{equation}
As it follows from the dependences presented on Fig.~\ref{fig2},
the estimate (\ref{L_e-a}) is in a good agreement with the results
of exact numerical calculations at small $a$.

We can approximately estimate the distance $L_e$ for a semi-infinite
medium in a similar manner assuming that the field of the reflected
wave in the wave zone and the field of the transmitted wave in the
medium are defined by Fresnel reflection and transmission
coefficients $r_F$ and $t_F$, respectively. Then for $s$-polarized
external wave after minor algebra we get
\begin{equation}
\label{Le-g}
\frac{L_e}{a}
=
\frac{1}{2\pi}
\ln
\left|
    \frac
    {8\pi^2 C t_F^s}
    {
      r_F^s
      \left(
           1-\frac{4\pi}{3}C
      \right)
    }
\right|
.
\end{equation}
In the case of the normal incidence
$r_F^s = -(n-1)/(n+1)$ and $t_F^s = 2/(n+1)$.
Taking into account the Lorentz-Lorenz formula (\ref{LL}), we obtain
\begin{equation}
\label{Le-si}
\frac{L_e}{a}
=
\frac{1}{2\pi}
\ln
\left[
    4\pi (n+1)
\right]
.
\end{equation}
From eq.(\ref{Le-si}) it folows that if the refractive index $n$
is in the range $1.1 \div 4.0$, then $L_e/a$ is in the range $0.5 \div 0.7$.
In fact $L_e/a$ has to be higher, because as it follows from the numerical
calculations in section 4
the local field inside the film in the first atomic layer is higher
than that in the ``bulk".

In the case of $p$-polarized external wave incident on a semi-infinite
medium the estimation of $L_e$ will be given by an equation which has
similar structure as (\ref{Le-g}) with $t_F^s$ and $r_F^s$ replaced by
$t_F^p$ and $r_F^p$, respectively. In this case near the Brewster angle
the reflection can be rather small but still noticeable. In such a regime
we get the values of $L_e/a$ which are somewhat bigger than $1$.
Therefore, our estimations show that the
propagating and evanescent harmonics can make comparable contributions
to the optical response of the medium at the distances from the surface
of the order of one lattice constant.

\subsubsection
{
Non-transversality of the total fields of reflected and
transmitted waves
}

\begin{figure}
\begin{center}
\leavevmode
\begin{tabular}{ll}
\epsfysize=6cm \epsffile{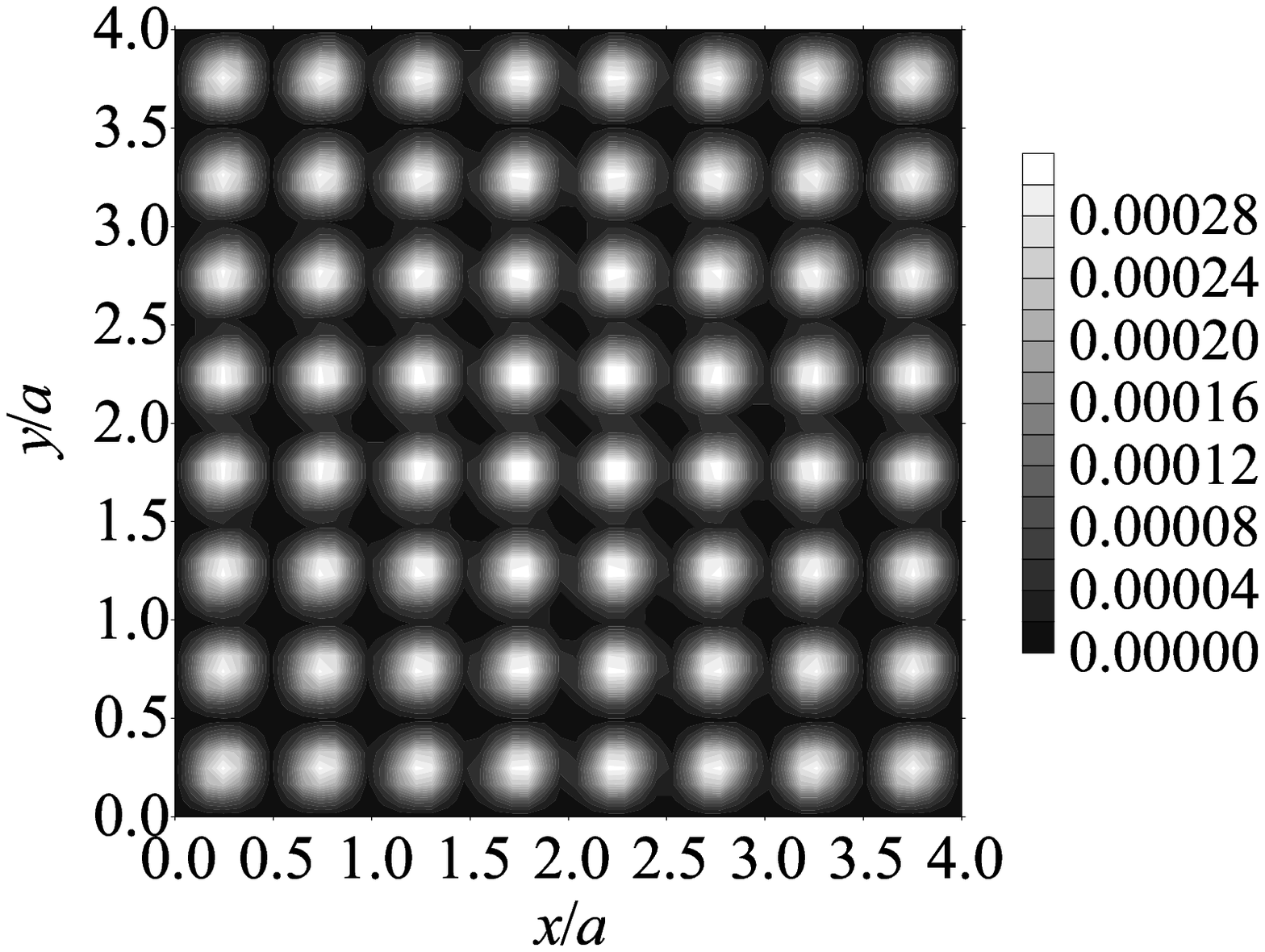} & \epsfysize=6cm \epsffile{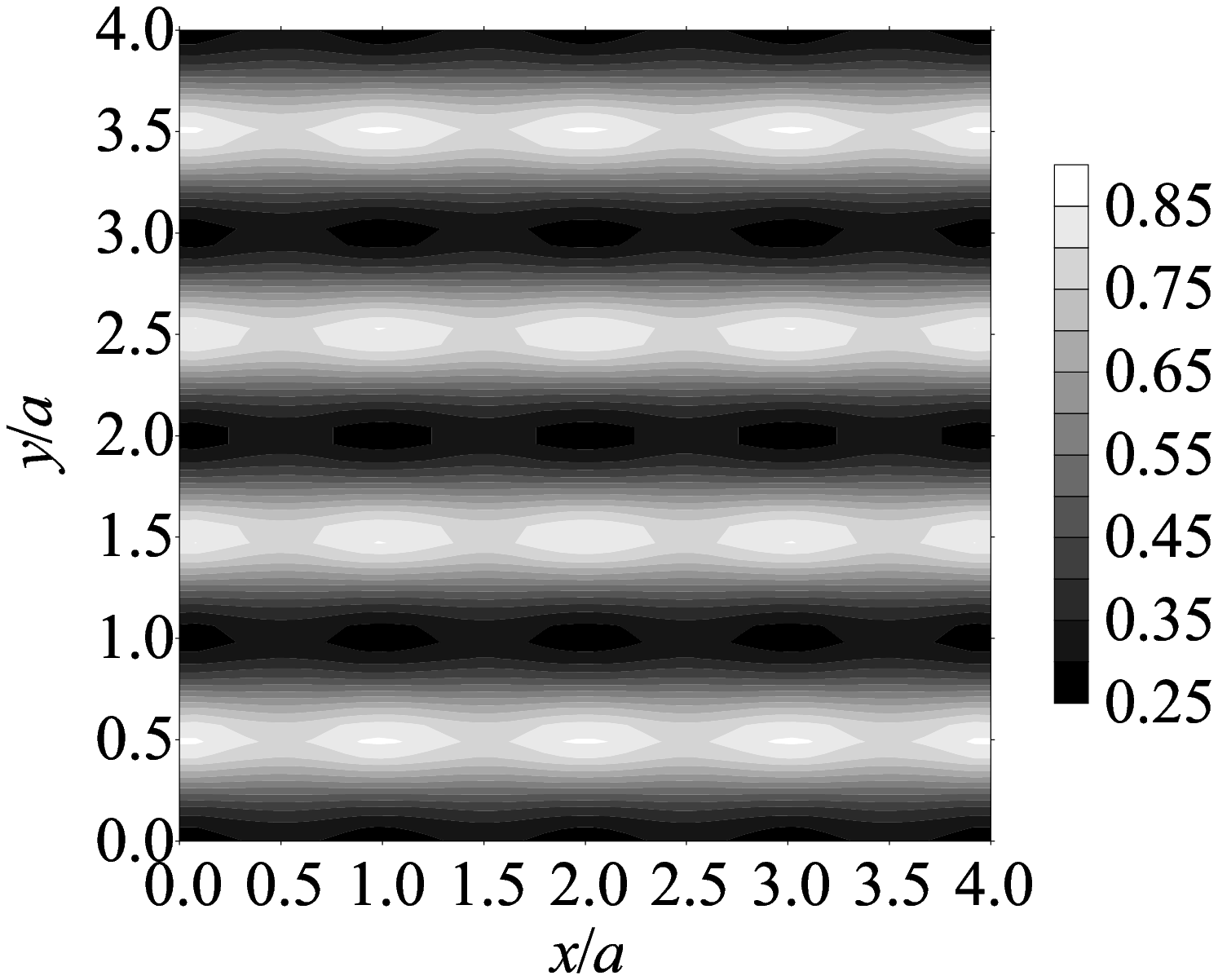}\\
\epsfysize=6cm \epsffile{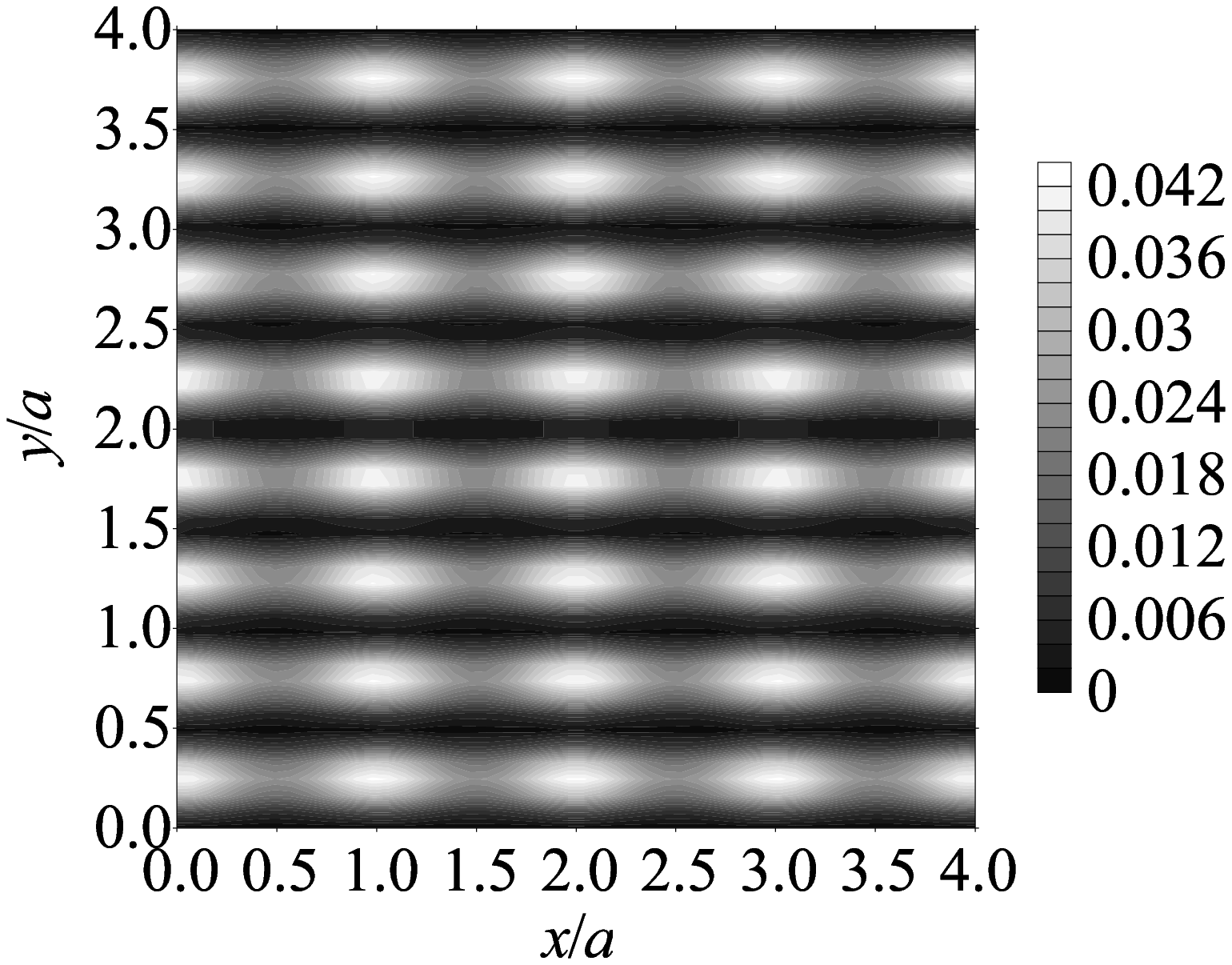} &
\end{tabular}
\end{center}
\caption
{
The ratio of the intensities of $x$- (a), $y$- (b) and $z$- (c)
components of the electromagnetic field on the probe
($I_x \sim |E_x|^2$, $I_y \sim |E_y|^2$, $I_z \sim |E_z|^2$)
to the intensity of the incident wave $I_0$ on the coordinates
$x$ and $y$.
The parameters are: the number of monolayers in the film
$N=20$, $C=0.2$, $k_0=0.01\ {\rm nm}^{-1}$, $a=0.5\ {\rm nm}$, $d=a$,
$\Theta_I=0^o$, $\Phi_I=0^o$. The intensity minima correspond
to the atomic positions.
In spite of the fact that the external wave is polarized along the
y axis, optical responds of the medium contains all three components.
}
\label{fig3}
\end{figure}

In distinction to the wave zone the polarization vector of the field
in the near zone has no a definite direction. Its magnitude and the
direction depend on all three spatial coordinates. Indeed, the polarization
vector of a certain evanescent harmonic in the r.h.s of eq.(\ref{32}) is
perpendicular to the wave vector ${\bf k}_{pq}$, which in its turn has
different orientations depending on $p$ and $q$. In addition, each
harmonic has its own decay coefficient $\kappa_{pq}$ and the behavior
of the harmonic along the surface is defined by its own vector
${\bf g}_{pq}^{\|}$. Carrying out the summation over all the harmonics
we get the properties of the polarization vector mentioned above. In
particular, in the case of the normal incidence of the external wave
with the polarization vector along the $y$ axis the field near
the film surface has non-vanishing $x$- and $z$-components
(Fig.~\ref{fig3}a,c).

For the purposes of the further analysis let's estimate the ratio
of the evanescent component of the reflected wave field to the
incident wave field. From the Airy formulae for the
$s$-polarization it follows that in the case of ultrathin film
the local field inside the film is $E_0\approx E_{0I}/[1-(4\pi /3)C]$.
The evanescent field at the distance of one lattice constant from the
surface is mainly determined by the harmonic $E_1$. For the square lattice
$E_1=-2(2\pi)^2 C \exp(-2\pi) E_0\approx 0.147 C/[1-(4\pi /3)C]E_{0I}$.
Even for $C=0.2$ the ratio $E_1/E_{0I}\approx 0.2$. Thus, the amplitude
of the evanescent wave at the distance from the surface of the order of
one lattice constant is usually much less than the incident field amplitude.
Therefore, one can conclude that in the field of the probe the main
contribution is made by the component of the field directed along the
polarization vector of the external wave. In particular, comparing
the intensities of $x$ and $z$-components of the field (Fig.~\ref{fig3}a,c)
with $y$-component (Fig.~\ref{fig3}b) one can notice that the
contribution of $x$- and $z$-components is small and the field behavior
near the surface is mainly determined by $y$-component.

One can come to one more conclusion regarding the direction of
the field polarization vector near the film surface. Under the
normal incidence of the external wave in the case when there are
no parallel shifts of the atomic planes relative to one another,
i.e.,
$\alpha_{1j}=\alpha_{2j}=0$, $j=\overline{1,N}$,
at the observation points above the atoms the
$z$-component of the reflected wave field vanishes (Fig.~\ref{fig3}).
Indeed, taking into account the properties of the tensor
$\hat f$ and summing in (\ref{field-l}) the harmonics with equal
$\kappa_{pq}$ ($\kappa_{pq}=\kappa_{-p,-q})$, we get that
$({\bf E}_j^0)_{x,y}$ are independent of $({\bf E}_j^0)_{z}$.
The $z$ components of the field inside the film satisfy to the system
of homogeneous equations which has only the trivial solution
$({\bf E}_j^0)_{z}\equiv 0$, $j=\overline{1,N}$. Then it follows
immediately from (\ref{32}) that at the observation points above the
atoms $({\bf E}_p)_z=0$.

\subsubsection
{
Dependence of field intensity on the longitudinal coordinates
$x$ and $y$
}

The electromagnetic field intensity is defined by its strength squared.
In a general case it can be calculated numerically. But in the case
of the normal incidence ($\Theta_I=0$) one can obtain rather simple
analytical expressions. Let's consider this case. From the plots on
Fig.~\ref{fig2} it follows that the component of the field, which is
a periodic function of longitudinal coordinates, is induced mainly
by the surface atoms. The bulk atoms produce only a constant background.
Thus, in order to analyze the behavior of the evanescent waves it is
enough to consider only one surface monolayer ($z_j=0$ in eq.(\ref{32})).
For the qualitative analysis of the field behavior near the film surface
it is enough to keep in eq.(\ref{32}) only the harmonics with
$(p,q)=(1,0)$, $(-1,0)$, $(0,1)$, $(0,-1)$
(we assume that the quantities ${g}_1$ and ${g}_2$ differ from one
another no more that two times). The amplitudes of all other harmonics
are at least one order of magnitude less~\cite{hex}. Omitting the
unimportant constant factor after an elementary algebra we obtain an
approximate expression for the field intensity at the probe position:
\begin{eqnarray}
\label{13}
I
&\approx&
\Biggl[
    E_{0I}+ {\rm Re} (E_R^{far})
\\
    & - &
    4\pi Ca
    \sum_{i=1,2} \frac{({\bf g}_i {\bf E}_{0I})}{E_{0I}}
    \left\{
    {\rm Re}({\bf g}_i {\bf E}_1^0)
    \cos\left[{\bf g}_i \left({\bf r}-{\bf r}_1^0\right)\right]
    -
    {g}_i {\rm Re}(E_{1z}^0)
    \sin\left[{\bf g}_i \left({\bf r}-{\bf r}_1^0\right)\right]
    \right\}
    \frac{\exp(-{g}_i z)}{{g}_i}
\Biggr]^2
,
 \nonumber
\end{eqnarray}
which is valid for the films of arbitrary thickness.
Here $E_1^0$ is an amplitude of the unperturbed part of the field
in the first (surface) monolayer. In fact, this expression is the
second power of the real part of the field component at the probe position,
directed along the polarization vector of the external field. The
imaginary part of this component and all other components are negligible.

From the expression (\ref{13}) one can see that in general the intensity
minima and maxima are shifted relative to the atomic positions. This is
due to the term which contains $E_{1z}^0$. Due to the fact that even in the
case of the normal incidence the dependence of the direction of the vector
${\bf E}^0_1$ on the direction of ${\bf E}_{0I}$ is rather complicated,
the location of the intensity minima and maxima depends on ${\bf E}_{0I}$
in a complicated manner as well. Let's consider the case when there are
no parallel shifts of the atomic planes relative to one another. As it was
shown above, in this case $E_{1z}^0=0$ and the expression (\ref{13})
becomes simpler. In this case the intensity minima are located exactly
at the positions of the atoms. Indeed, at these points
$
\cos[{\bf g}_1 ({\bf r}-{\bf r}_1^0)]
=
\cos[{\bf g}_2 ({\bf r}-{\bf r}_1^0)]=1
$.
Thus, formula (\ref{13}) explains the contrast reversal, which has been
obtained in Refs.~\cite{Keller, Girard} by means of numerical calculations.

\begin{figure}
\begin{center}
\leavevmode
\begin{tabular}{cc}
\epsfysize=6cm \epsffile{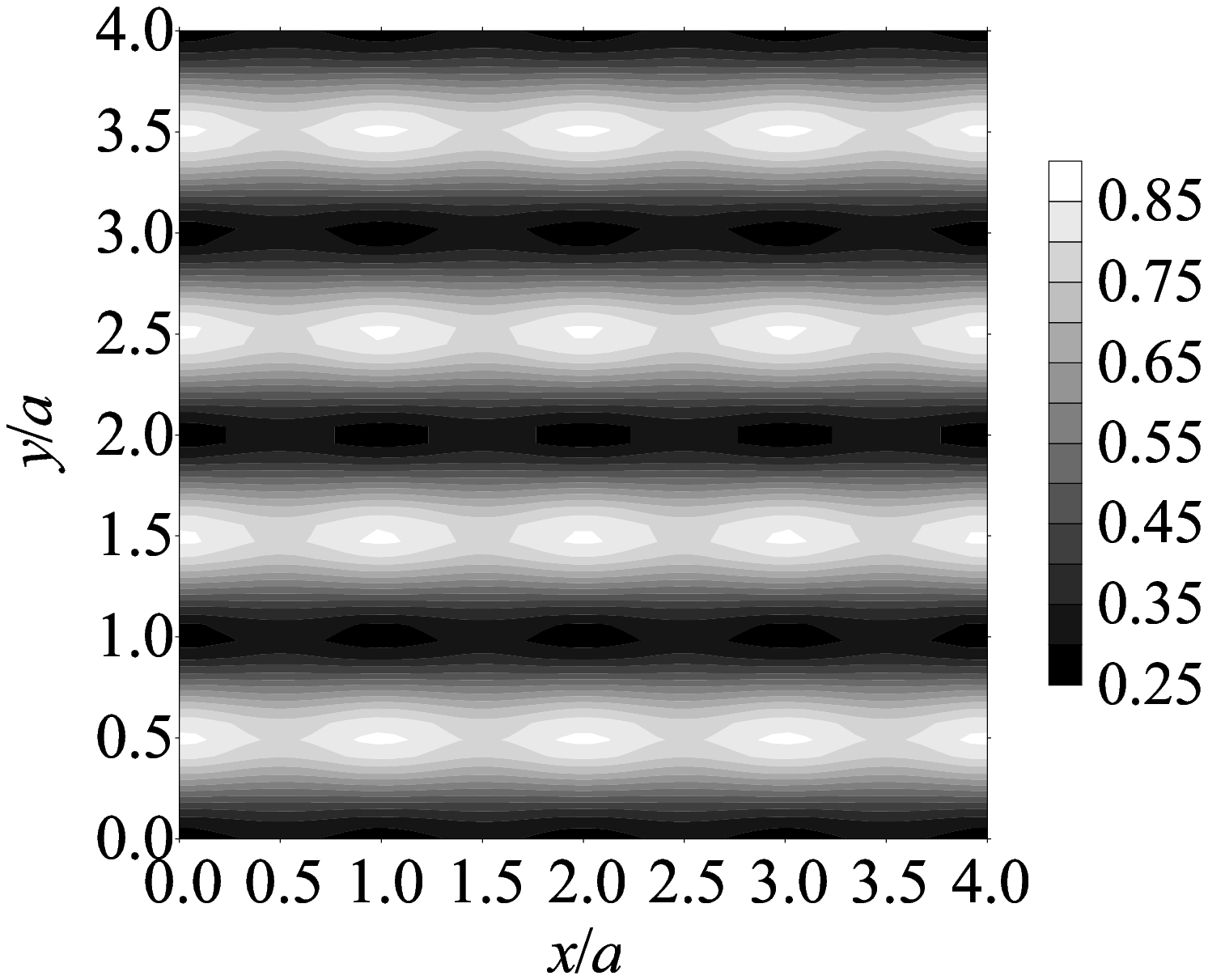} & \epsfysize=6cm \epsffile{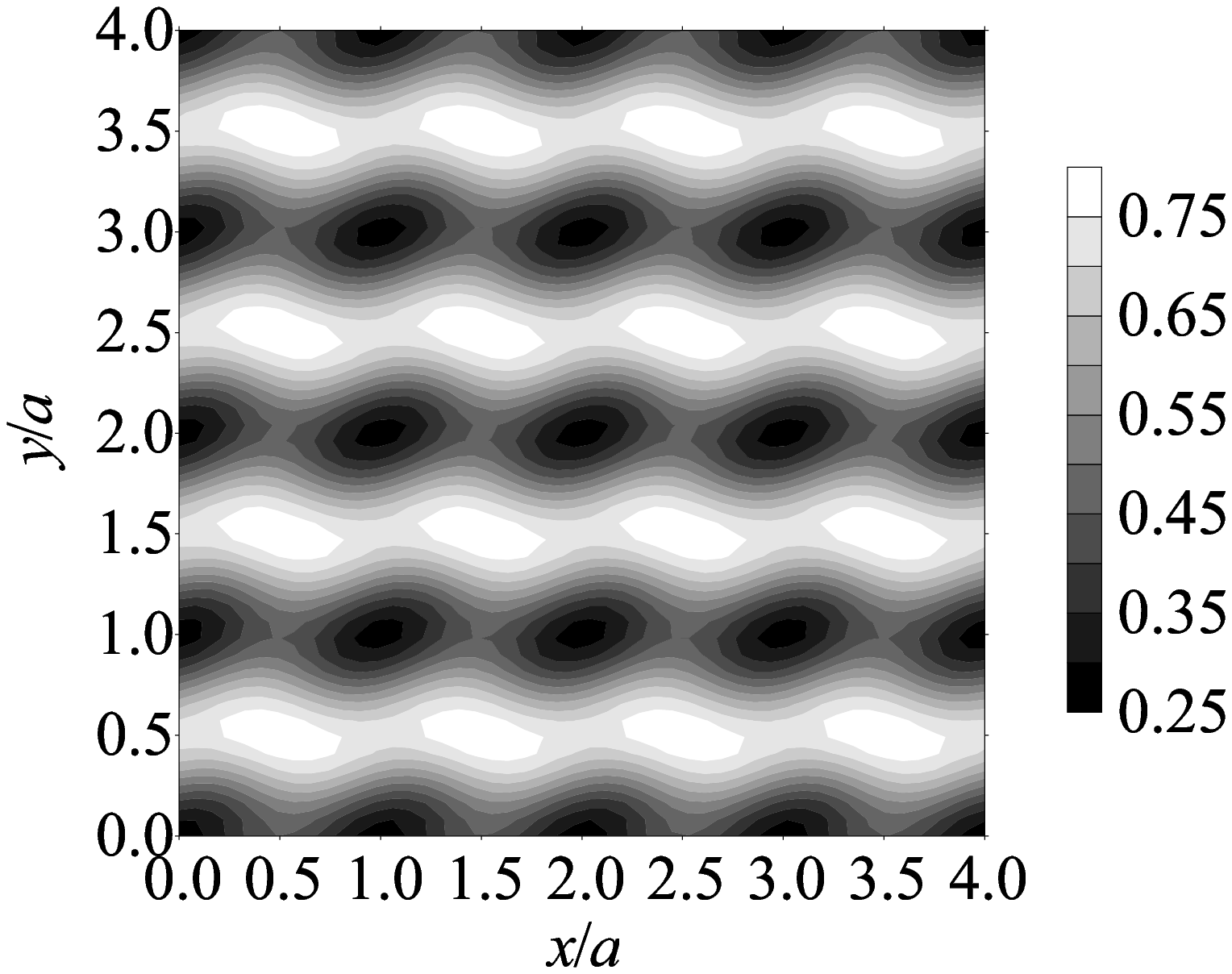}\\
\epsfysize=6cm \epsffile{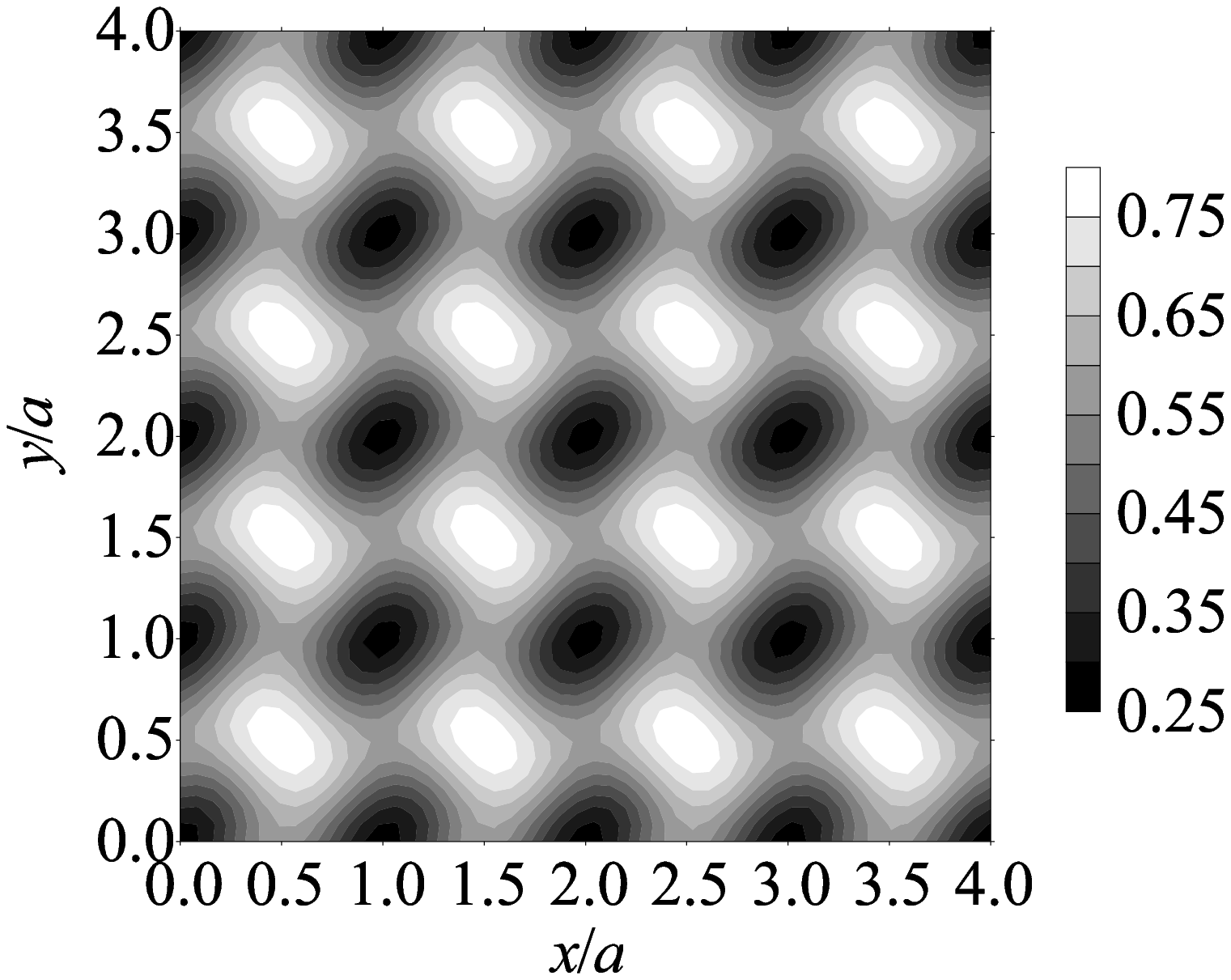} &
\end{tabular}
\end{center}
\caption{
The ratio of the electromagnetic field intensity on the probe
$I$ to the incident wave intensity $I_0$.
The parameters are: $N=20$, $C=0.2$, $k_0 = 0.01\ {\rm nm}^{-1}$,
$a=0.5\ {\rm nm}$, $d=a$, $\Theta_I=0^o$;
$\Phi_I=0^o$ (a), $30^o$ (b), $45^o$ (c).
The electromagnetic field intensity is a periodic function
of the coordinates $x$ and $y$, and the period is equal to the lattice
constant $a$. The intensity minima correspond to the atomic positions.
}
\label{fig4}
\end{figure}

In analogous manner one could assume that the intensity maxima are
located in the centers of the elementary cells, where the cosines are
equal to $-1$. However, this is not exactly the case as one can see
on Fig.~\ref{fig4}. 
With the change of the mutual orientation of the
polarization vector of the external wave and the lattice elementary
translations vectors the location of the intensity maxima changes as well.
The reason is the following. Let the polarization vector of the external
wave is perpendicular to one of the vectors of the reciprocal lattice,
for instance, $({\bf E}_{0I} {\bf g}_1)=0$.
The field behavior along the direction ${\bf a}_1$ in this case is
determined by higher order harmonics with the smallest decay coefficient
and with non-vanishing component along the polarization vector of the
external wave. Obviously, these are harmonics with $|p|=|q|=1$.
In eq.(\ref{13}) in the square brackets one has to add one more term
  $$
-4 \pi Ca
\frac{({\bf g}_{11}^{min}{\bf E}_{0I})}{E_{0I}}
{\rm Re}({\bf g}_{11}^{min} {\bf E}_1^0)
\cos
\left[
    {\bf g}_{11}^{min} \left({\bf r}-{\bf r}_1^0\right)
\right]
\frac{\exp (-{g}_{11}^{min} z)}{{g}_{11}^{min}}
\;,
  $$
where ${\bf g}_{11}^{min}$ is one of the vectors ${\bf g}_{pq}$
with $|p|=|q|=1$, which has the minimal length. At the observation points
above the atoms (${\bf r}^\|={\bf r}_{a_1}^\|$) this term is minimal,
i.e., the intensity minima are still above the atoms, even provided that
$({\bf E}_{0I}{\bf g}_1)=0$ or $({\bf E}_1^0 {\bf g}_1)=0$. The locations
of the intensity maxima can be found from the condition
$({\bf g}_{11}^{min} {\bf r})=\pi\pm 2\pi k$, where $k$ is an integer.
This condition fulfills not in the centers of the elementary cells,
like in the case when the directions of the polarization vector of the
external wave ${\bf E}_{0I}$ and the local field inside the film
${\bf E}_1^0$ does not coincide with any of the translation vectors
of the lattice (Fig.~\ref{fig4}b,c), but at the points
${\bf r}^\|={\bf r}_{a_1}^\|+{\bf a}_1/2$ or
${\bf r}^\|={\bf r}_{a_1}^\|+{\bf a}_2/2$ (Fig.~\ref{fig4}a).
Thus, in distinction to the intensity minima locations, which are always
above the atoms, the locations of the intensity maxima depend on the
mutual orientation of the field polarization and the translation vectors
of the lattice.

Formula (\ref{13}) allows one to investigate the image contrast, which
is defined here as a maximal difference of the light intensity along
a certain direction. In Refs.~\cite{Xiao,Keller,Reddick} it was point out
that the field component parallel to the scan direction
displays a better contrast than that perpendicular to this direction.
Let's consider this problem in more details. The image contrast
in the direction ${\bf a}_i$ is determined by the mutual orientation
of the vectors ${\bf E}_{0I}$, ${\bf E}_1^0$, and ${\bf g}_i$.
The maximal contrast is reached at some angle between
${\bf E}_{0I}$ and ${\bf g}_i$ when the coefficient
$({\bf g}_i{\bf E}_{0I}) {\rm Re}({\bf g}_i {\bf E}_1^0)$
takes the largest value. This angle depends in general on the symmetry
of the atomic distribution at the surface and on the atomic polarizability.
We are going to come back to this issue in our subsequent publications.

The minimal contrast along the direction ${\bf a}_1$ for the lines
which pass through the atoms is achieved when the polarization vector
of the external wave or the field vector in the film is perpendicular
to the vector of the reciprocal lattice ${\bf g}_1$
(parallel to the vector of the real lattice ${\bf a}_2$) (\ref{13}).
Performing numerical calculations, one can show that in general the
conditions $({\bf g}_i {\bf E}_{0I})=0$ and $({\bf g}_i {\bf E}_1^0)=0$
are satisfied at different directions of the polarization vector of the
external wave. Thus, in general there exist two angles at which contrast
minima can be observed. The angle between the two directions of the
polarization vector of the external field allows one to get the information
about the local field inside the film and, therefore, about the atomic
polarizability.

Another situation occurs for the contrast along the lines which pass through
the centers of the elementary cells. In this case the contrast minima
are determined by the condition of mutual compensation of the contributions
from the harmonics with decay coefficients ${g}_1$ and ${g}_{11}^{min}$.
The compensation occurs at some angle $\phi_c$ between the polarization
vector of the field inside the film and the vector ${\bf g}_1$,
which depends on the symmetry of the atomic distribution at the surface.
The angle $\phi_c$ strongly depends on the distance from the surface.

\begin{figure}
\epsfxsize=8cm \centerline{\epsffile{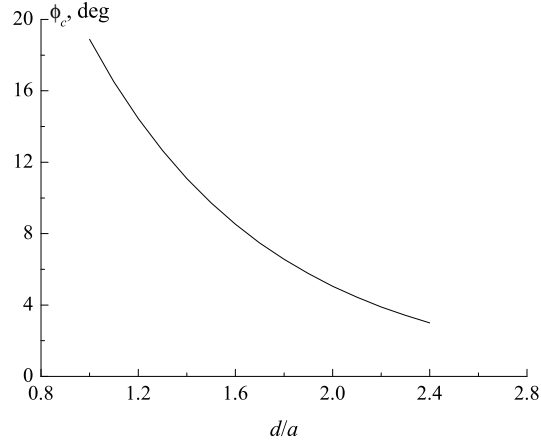}}
\caption{
The angle between the minimal contrast direction and the polarization
vector of the external wave $\phi_c$ as a function of the distance
to the film surface $d$.
The calculations are performed for the cubic lattice,
$\Theta_I=0^o$.
}\label{fig5}
\end{figure}

Let's consider the simplest case of the cubic lattice. Under the normal
incidence of the external wave the direction of the polarization vector
of the field inside the film coincides with that of the external field.
In this case there are two directions of the polarization vector of the
external wave, for which one can observe the contrast minima along
the lines which pass through the atoms. This corresponds to the situations
when the polarization vector coincides with ${\bf a}_1$ or ${\bf a}_2$.
For the cubic lattice one can get an explicit expression for the angle
$\phi_c$ at which one observes the minimal contrast along the lines which
pass through the centers of the cells. The dependence of the angle
$\phi_c$ on the distance between the observation point and the surface
is given by the approximate formula
\begin{equation}
\label{28}
\sin\phi_c
=
2^{1/4}
\exp
\left[
    - \pi (\sqrt{2}-1) d/a
\right]
\;,
\end{equation}
which is valid only for $d \ge a$. The dependence
$\phi_c(d)$ for the cubic lattice is shown on Fig.~\ref{fig5}.
As one can see on the figure, the angle $\phi_c$ can take the values
of dozens of degrees at the distance from the surface of the order of
one lattice constant. The magnitude of the angle remains essential
at the distances of several lattice constants.

Thus, changing the direction of the polarization vector of the external
wave one can manipulate the image contrast. One can determine the atomic
positions and the symmetry of the crystal lattice detecting the intensity
minima. Measuring the angle between the directions of the polarization
vector of the external wave, at which the minimal image contrast is reached,
one can get the information about the local fields inside the film.
Having on hand this information and measuring the angle $\phi_c$
between the translation vectors of the lattice and the polarization vector
of the external wave one can determine the distance between the probe
and the surface.

The analysis based on eq.(\ref{F}) allows one to say that formula
(\ref{13}) remains valid in the case of arbitrary incident angle
of $s$-polarized wave as well. One has to note that in the
near-field optical measurements it is more useful to work with
$p$-polarized waves~\cite{Keller}. In the case of arbitrary incidence
angle of $p$-polarized wave the distribution of the field in the
near zone is essentially different. This has been shown
in Ref.~\cite{Miyazaki} on the basis of numerical calculations.
We intend to give a detailed description of the field behavior in
the near zone at arbitrary incident angles in our subsequent
publications.

\section{Conlcusions}

We have solved a boundary problem of the linear
classical optics devoted to the investigation of the electromagnetic
field behavior near the surface of a dielectric medium taking into
account its discrete structure. The main attention has been paid to
the investigation of the near-zone optical properties of dielectrics.
It has been shown that at the distances
from the surface less than two lattice constants the behavior of the
reflected and transmitted waves is entirely different from what we have
in the wave zone. The intensity distribution in the near zone allows one
to determine the atomic positions at the surface.

We have shown that at the distances from the surface larger than the
interatomic distance in the film the probe which measures the field
does not significantly influence the field distribution in the film.
In the present paper we have suggested a method of the optical control
of the distance between the probe and the surface. The method is based
on the analysis of the near-field distribution at different directions
of the polarization vector of the external radiation.

At first glance the system treated in our paper (ultrathin dielectric
film without any substrate) seems to be unrealistic. However, it can be used
for the investigation of the near-zone optical properties of the films
of arbitrary thicknesses. As we have shown the near-zone optical response
of a material is defined mainly by the evanescent harmonics, which play
an important role only within few interatomic distances near the surface,
because they decay very rapidly. This means that the ``bulk" monolayers
influence significantly only the wave-zone optical properties and does not
change much the ratio between propagating and evanescent harmonics.

Our results can be useful not only for
the development of the ultrahigh resolution near-field microscopy,
but also for the investigation of the optical properties of photonic
crystals in the long wavelength approximation.

The results obtained in the present paper are valid not only in
the case of an ideal crystal lattice, but also when the medium
has only short-range order as in the case of real surfaces.
Indeed, the methods based on the Fourier transform and the
Lorentz method give analogous results. At the same time only the
short-range order is important for the Lorentz method and the
magnitude of the dipole field is mainly determined by the atoms
located at the distance from the observation point of the order
of $2\div 3$ lattice constants. The atomic
distribution outside of this region does not practically
influence the field.

\section*{Acknowledgement}
This work has been supported in part by the Russian federal program
``Integration" (grant A 0066).
One of the authors (K.V.K.) is grateful to
Deutsche Forschungs\-gemeinschaft and Alexander-von-Humboldt Stiftung
for financial support.

\newpage



\begin{thebibliography}{99}
\bibitem{hex}
For the description of the field near hexagonal lattice it is
necessary to take into account 6 harmonics. The detailed analysis
of the behavior of the near field for this case can be found in
Ref.~\cite{Miyazaki}.
\end{thebibliography}
\end{document}